\documentclass[twocolumn]{autart}    

\usepackage{graphicx}          

\usepackage{amsmath,amssymb,amsfonts}
\usepackage{amsmath}
\usepackage{amssymb}
\usepackage{bm}
\usepackage{dsfont}
\usepackage{bbm}
\usepackage{mathtools}
\usepackage{commath}
\usepackage{yhmath}
\usepackage{mathrsfs}
\usepackage{algorithm}
\usepackage{algorithmicx}
\usepackage{enumerate}
\usepackage[noend]{algpseudocode}
\usepackage[compress]{cite}
\usepackage{hyperref}
\hypersetup{colorlinks=true, linkcolor=blue, breaklinks=true, urlcolor=blue, citecolor=blue}
\usepackage{natbib}
\usepackage{pgfplots}

\usepackage{caption}

\newtheorem{defi}{\textbf{Definition}}
\newtheorem{thom}{\textbf{Theorem}}

\newtheorem{rek}{\textbf{Remark}}

\newtheorem{lema}{\textbf{Lemma}}

\newcommand{\defiref}[1]{Definition \ref{#1}}
\newcommand{\thomref}[1]{Theorem~\ref{#1}}

\newcommand{\rekref}[1]{Remark~\ref{#1}}

\newcommand{\algoref}[1]{Algorithm \ref{#1}}

\newcommand{\lemaref}[1]{Lemma \ref{#1}}

\newcommand{\pnum}{N_p}
\newcommand{\enum}{N_e}
\newcommand{\pteam}{\mathscr{P}}
\newcommand{\eteam}{\mathscr{E}}

\newcommand{\goal}{\Omega_{\rm goal}}
\newcommand{\play}{\Omega_{\rm play}}
\newcommand{\targetline}{\mathcal{T}}

\newcommand{\interior}{\mathrm{int}}

\newcommand{\revise}{\color{black}}

\date{}

\begin{document}

\begin{frontmatter}

\title{Multiplayer Homicidal Chauffeur Reach-Avoid Games: A Pursuit Enclosure Function Approach} 

\thanks[footnoteinfo]{The work of R. Yan was partly completed at Tsinghua University and also funded in part by the ERC under the European Union’s Horizon 2020 research and innovation programme (FUN2MODEL, grant agreement No.~834115). The work of X. Duan was sponsored by Shanghai Pujiang Program under grant 22PJ1404900. The work of R. Zou, X. He and Z. Shi was supported by the Science and Technology Innovation 2030-Key Project of "New Generation Artificial Intelligence" under Grant 2020AAA0108200.}

\author[Oxford,Tsinghua]{Rui Yan}\ead{rui.yan@cs.ox.ac.uk} ,
\author[SJTU]{Xiaoming Duan}\ead{xduan@sjtu.edu.cn},
\author[Tsinghua]{Rui Zou}\ead{zr20@mails.tsinghua.edu.cn},
\author[Tsinghua]{Xin He}\ead{hex20@mails.tsinghua.edu.cn},
\author[Tsinghua]{Zongying Shi}\ead{szy@mail.tsinghua.edu.cn},
\author[UCSB]{Francesco Bullo}\ead{bullo@ucsb.edu} 

\address[Oxford]{Department of Computer Science, University of Oxford, Oxford OX1 3QD, UK}
\address[Tsinghua]{Department of Automation, Tsinghua University, Beijing 100084, China}
\address[SJTU]{Department of Automation, Shanghai Jiao Tong University, Shanghai 200240, China}
\address[UCSB]{Center of Control, Dynamical Systems and Computation, University of California, Santa Barbara, USA}




\begin{abstract}
This paper presents a multiplayer Homicidal Chauffeur reach-avoid differential game, which involves Dubins-car pursuers and simple-motion evaders. The goal of the pursuers is to cooperatively protect a planar convex region from the evaders, who strive to reach the region. We propose a cooperative strategy for the pursuers based on subgames
for multiple pursuers against one evader and optimal task allocation. We introduce pursuit enclosure functions (PEFs) and propose a new enclosure region pursuit (ERP) winning approach that supports the forward analysis for the strategy synthesis in the subgames. 
We show that if a pursuit coalition is able to defend the region against an evader under the ERP winning, then no more than two pursuers in the coalition are necessarily needed. We also propose a steer-to-ERP approach to certify the ERP winning and synthesize the ERP winning strategy. To implement the strategy, we introduce a positional PEF 
and provide the necessary parameters, states, and strategies that ensure the ERP winning for both one pursuer and two pursuers against one evader. Additionally, we formulate a binary integer program using the subgame outcomes to maximize the captured evaders in the ERP winning for the pursuit task allocation. Finally, we propose a multiplayer receding-horizon strategy where the ERP winnings are checked in each horizon, the task is allocated, and the strategies of the pursuers are determined. Numerical examples are provided to illustrate the results.
\end{abstract}

\end{frontmatter}

\section{Introduction}

\emph{Problem motivation and description:} 
Multi-robot systems with adversarial goals, failures or improper uses,
could pose significant threat to safety-critical infrastructure, such as airports and military facilities. We consider a planar multiplayer Homicidal Chauffeur reach-avoid differential game. In this game, multiple Dubins-car pursuers are used to protect a convex region against a number of malicious simple-motion evaders. To capture the most number of evaders, we propose a receding-horizon strategy for the pursuers based on subgames for multiple pursuers against one evader and optimal task allocation.

\emph{Literature review:} The classical Homicidal Chauffeur differential game first proposed by \cite{RI:65}, is an attractive pursuit-evasion game between a Dubins-car pursuer and a simple-motion evader, and has a long research history \citep{VSP-VLT:11}. The strategies for the players in this game are quite complex mainly due to the non-linearity of the Dubins-car dynamics, and it took a long time to finally obtain its complete solution
by \cite{AWM:71}. Since then, many interesting variants have been proposed, including surveillance-evasion objectives \citep{JL-JVB:75,JL-GJO-79}, the stochastic version \citep{MP-YY:81} and suicidal pedestrians \citep{IE-PT-MP:15}. Numerical investigation and multiplayer extensions can be found in \citep{SDB-FB-JPH:09,IMM:02,MF:06}.

Recently, reach-avoid differential games \citep{KM-JL:11,ZZ-RT-HH-CJT:12,RY-RD-XD-ZS-YZ:23}, also known as two-target differential games \citep{WMG-MP:81,PC:96}, perimeter defense games \citep{DS-VK:20}, or target guarding problems \citep{HF-HHL:23,JM-NK-BB:20}, have received considerable research attention due to the wide safety and security applications in antagonistic multiplayer systems. These games consider a scenario where a group of pursuers (or defenders) are tasked to protect a critical region from a group of evaders (or attackers). Due to the high-dimensional continuous joint action and state spaces, as well as complex cooperations and competitions among players, the current literature either focuses on one pursuer against one evader \citep{PC:96,KM-JL:11,IMM-AMB-CJT:05,JM-NK-BB:20,ZZ-RT-HH-CJT:12}, or solves a multiplayer game suboptimally by decomposing it into many subgames with a few players and piecing them together through task allocation \citep{RY-XD-ZS-YZ-FB:19,RY-RD-XD-ZS-YZ:23,RY-ZS-YZ:20-1,RY-ZS-YZ:19-2,YL-EB:22,MC-ZZ-CJT:17,DS-VK:18,DS-VK:20,EG-DWS-AVM-MP:20}. Three common approaches have been developed to determine the game winners or compute strategies for the players. The popular Hamilton-Jacobi-Isaacs (HJI) method using level sets relies on gridding over the state space and is ideal for low-dimensional systems \citep{KM-JL:11,ZZ-RT-HH-CJT:12,IMM-AMB-CJT:05,IMM:02}. The classical characteristic method \citep{AVM-MP-DS-ZF:22,EG-DWC-MP:20} involves integrating backward from non-unique terminal conditions (capture or entry into the protected region), and may generate complicated singular surfaces when different backward trajectories meet. The geometric method employs geometric concepts, such as Voronoi diagram \citep{RY-ZS-YZ:19,ZZ-WZ-JD-HH-DMS-CJT:16}, Apollonius circle \citep{RY-ZS-YZ:20-2,RY-ZS-YZ:19,JM-NK-BB:20}, function-based evasion space \citep{RY-XD-ZS-YZ-FB:19}, and dominance region \citep{DWO-PTK-ARG:16}, for both qualitative and quantitative analysis, and have been proved powerful especially in the case of simple-motion players.

However, there are limited works on Homicidal Chauffeur reach-avoid differential games which integrate the Homicidal Chauffeur dynamics with reach-avoid competition goals. Most of the research in reach-avoid games focuses on either
complex dynamical models (e.g., Dubins car) with numerical methods, or simple
dynamical models with analytical methods. HJI reachability has been applied to one-pursuer-one-evader cases \citep{MC-SB-JFF-CJT:19}, but would suffer from high computational burden for multiplayer games.  The most relevant work is \cite{RY-RD-HL-WZ-ZS-YZ:23} in which a feedback strategy is proposed with the guaranteed capture of an evader by a pursuer, provided that some  conditions on initial configurations are satisfied. In \cite{RY-RD-HL-WZ-ZS-YZ:23}, it mainly considers one pursuer against one evader for a protected region with infinite area and a linear boundary, and constructs pursuit strategies based on the Apollonius circle. 

\emph{Contributions:} We propose a receding-horizon cooperative pursuit strategy for multiplayer Homicidal Chauffeur reach-avoid differential games, with efficient computation and guaranteed winning performance. The main contributions are as follows:
\begin{enumerate}
    \item For the subgame with multiple pursuers (a pursuit coalition) and one evader, we introduce pursuit enclosure functions (PEFs) and then propose a new enclosure region pursuit (ERP) winning approach. The ERP winning and its strategies can be computed through the forward analysis instead of the backward reachability from the terminal conditions that generally involves solving HJI equations.

    \item 
    We prove that under the ERP winning, if a pursuit coalition is able to defend against an evader, then at most two pursuers in the coalition are needed. This largely simplifies the pursuit strategies, as only one-pursuer and two-pursuer coalitions are needed to ensure the win. 

    \item We propose a steer-to-ERP approach in which an optimization problem is solved, to generate new ERP winning states and synthesize the corresponding ERP winning strategies, based on the known ERP winning states.

    \item To implement the strategy, a positional PEF based on players' current positions is introduced. Parameters, states and strategies that can ensure the ERP winning are presented, for the cases of one pursuer and two pursuers against one evader. Finally, a multiplayer receding-horizon strategy is proposed such that in each horizon, the number of captured evaders in the ERP winning is maximized.
\end{enumerate}

\emph{Paper organization:} We introduce Homicidal Chauffeur reach-avoid differential games in Section \ref{sec:problem}. The ERP winning is proposed followed by a coalition reduction in Section \ref{sec:ERP-reduction}. In Section \ref{sec:sink-winning-subspace}, we propose a steer-to-ERP approach to generate new ERP winning states. Section \ref{sec:positional-PEF} introduces a positional PEF and presents the conditions to ensure the ERP winning. In Section \ref{sec:taks-allocation}, a multiplayer strategy is proposed. Numerical results are presented in Section \ref{sec:simulation} and we conclude the paper in Section \ref{sec:conclusion}.

\emph{Notation:} Let $\mathbb{R}$, $\mathbb{R}_{>0}$ and $\mathbb{R}_{\ge0}$ be the set of reals, positive reals and nonnegative reals, respectively. Let $\mathbb{R}^n$ be the set of $n$-dimensional real column vectors and $\enVert[0]{\cdot}_2$ be the Euclidean norm. All vectors in this paper are column vectors and $\bm{x}^\top$ denotes the transpose of a vector $\bm{x} \in \mathbb{R}^n$. Let $\bm{0}$ denote the zero vector whose dimension will be  clear from the context. 
Denote the unit disk in $\mathbb{R}^n$ by $\mathbb{S}^{n}$, i.e., $\mathbb{S}^{n}=\{ \bm{u}\in\mathbb{R}^n\, |\, \|\bm{u}\|_2 \leq 1 \}$. 
The distance between two points $\bm{x}\in\mathbb{R}^2$ and $\bm{y}\in\mathbb{R}^2$ is defined by $d(\bm{x}, \bm{y})=\enVert[0]{\bm{x}-\bm{y}}_2$. The distance between a point $\bm{x}\in\mathbb{R}^2$ and a non-empty set $\mathcal{M}\subset\mathbb{R}^2$ is defined by $d(\bm{x}, \mathcal{M})=\inf_{\bm{y}\in\mathcal{M}}\enVert[0]{\bm{x}-\bm{y}}_2$.
The distance between two non-empty sets $\mathcal{M}_1$ and $\mathcal{M}_2$ is defined by $d(\mathcal{M}_1,\mathcal{M}_2)=\inf_{\bm{x}\in\mathcal{M}_1,\bm{y}\in\mathcal{M}_2}\enVert[0]{\bm{x}-\bm{y}}_2$.
For a vector $\bm{x} = [x, y]^\top \in \mathbb{R}^2$, let $\bm{x}^\circ = [y, - x]^\top$ be the vector obtained by rotating $\bm{x}$ in a clockwise direction by $\pi /2$. For a finite set $S$, we denote by $|S|$ the cardinality of $S$. Further notations are provided in Table 1, which will be explained in more detail later.

 \begin{table*}
    \centering 
    \captionsetup{labelformat=empty}
    \begin{tabular}{ p{.17\linewidth}  p{.28\linewidth} p{.1\linewidth}  p{.33\linewidth} } 
       \multicolumn{4}{ c }{Table 1. Notation Table} \\
        \hline
        Symbol & Description & Symbol & Description\\
        \hline
    $X_{ij} = (\bm{x}_{P_i}, \theta_{P_i}, \bm{x}_{E_j})$ & state of $P_i$ and $E_j$ & $\mathcal{S}_{ij}$ & set of $X_{ij}$ when $E_j$ is not captured by $P_i$\\
        \hline
    $X_c = \{(\bm{x}_{P_i}, \theta_{P_i})\}_{i \in c}$ & state of pursuit coalition $P_c$ & $f_c = \{f_i\}_{i \in c}$ & set of PEFs for pursuit coalition $P_c$\\
    \hline
    $X_{cj}=(X_c,\bm{x}_{E_j})$ & state of $P_c$ and $E_j$ & $\mathcal{S}_{cj}$ & set of $X_{cj}$ when $E_j$ is not captured by $P_c$\\
    \hline
    $\mathbb{E}(X_{cj}; f_c)$ & enclosure region for $P_c$ and $E_j$ & $\varrho(X_{cj}; f_c)$ & safe distance for $P_c$ and $E_j$ \\
    \hline 
    $X_{cj}^{t}$ & state after $t$ starting from $X_{cj}$ & $\mathcal{P}(X_{cj})$ & convex program for computing $\varrho(X_{cj}; f_c)$ \\
    \hline
     $f_i^{\textup{ps}}, f_c^{\textup{ps}}$ & positional PEF(s) & $\mathcal{X}_{cj}^{\textup{ps}} $ & an initial set of ERP winning states \\
    \hline
    $\varrho(X_{ij}; f_i^{\textup{ps}}), \varrho(X_{cj}; f_c^{\textup{ps}})$ & positional safe distance  & $\mathcal{P}^{\textup{ps}}(X_{cj})$ & convex program for computing $\varrho(X_{cj}; f_c^{\textup{ps}})$\\
    \hline
    \end{tabular}

    \label{tab:notation}
    \end{table*}

\section{Problem Statement}\label{sec:problem}

\begin{figure}
    \centering
    \includegraphics[width=0.9\hsize]{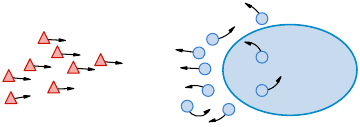}
    \put(-105,55){\footnotesize$\pteam$}
    \put(-160,50){\footnotesize$\eteam$}
    \put(-35,30){\footnotesize$\goal$}
    \put(-35,63){\footnotesize$\targetline$}
    \put(-150,5){\footnotesize$\play$}
    \caption{Multiplayer Homicidal Chauffeur reach-avoid differential games, where a group of (red) simple-motion evaders $\eteam$, starting from a play region $\play$, aim to enter a goal region $\goal$ protected by multiple  (blue) car-model pursuers $\pteam$, and $\targetline$ is the boundary curve between $\play$ and $\goal$.}
    \label{fig:game-illustration}
\end{figure}
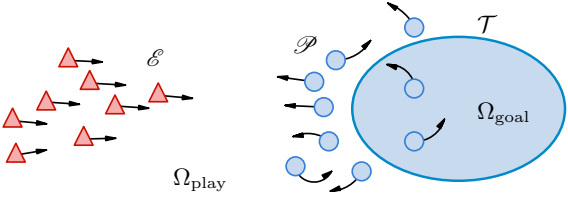

\subsection{Homicidal Chauffeur reach-avoid games}
Consider a Homicidal Chauffeur reach-avoid differential game in an obstacle-free plane between $\pnum$ pursuers $\pteam=\{P_1,\dots,P_{\pnum}\}$ and $\enum$ evaders $\eteam=\{E_1,\dots,E_{\enum}\}$.
Each pursuer $P_i \in \pteam$ is a Dubins car:
\begin{equation}\label{eq:pursuer_car}
    \begin{aligned}
        \dot{x}_{P_i}&=v_{P_i}\cos\theta_{P_i},& x_{P_i}(0)&=x_{P_i}^0,\\
        \dot{y}_{P_i}&=v_{P_i}\sin\theta_{P_i},&y_{P_i}(0)&=y_{P_i}^0,\\
        \dot{\theta}_{P_i}&=v_{P_i}u_{P_i}/\kappa_i,&\theta_{P_i}(0)&=\theta_{P_i}^0,
    \end{aligned}
\end{equation}
where $\bm{x}_{P_i}=[x_{P_i},y_{P_i}]^\top \in \mathbb{R}^2$, $\theta_{P_i}\in \mathbb{R}$ and $u_{P_i}$ are  $P_i$'s position, heading and control input, respectively, and $v_{P_i}, \kappa_i \in \mathbb{R}_{>0}$ are $P_i$'s maximum speed and minimum turning radius, respectively. Assume that $u_{P_i}$ belongs to $\mathbb{U}_P = \{ u : [0, \infty) \rightarrow \mathbb{S}^1 \,|\, u \textup{ is piecewise smooth}\}$. The initial position and heading are $\bm{x}_{P_i}^0=[x_{P_i}^0,y_{P_i}^0]^\top \in \mathbb{R}^2$ and $\theta_{P_i}^0 \in \mathbb{R}$, respectively. Each evader $E_j \in \eteam$
has a simple motion:
\begin{equation}\label{eq:evader_simple}
    \begin{aligned}
        \dot{x}_{E_j}&=v_{E_j}u_{E_j}^x,&x_{E_j}(0)&=x_{E_j}^0,\\ 
        \dot{y}_{E_j}&=v_{E_j}u_{E_j}^y,&y_{E_j}(0)&=y_{E_j}^0,
    \end{aligned}
\end{equation}
where $\bm{x}_{E_j} = [x_{E_j}, y_{E_j}]^\top \in \mathbb{R}^2$ and $\bm{u}_{E_j} = [u_{E_j}^x, u_{E_j}^y] ^ \top$ are $E_j$'s position and control input, respectively, $v_{E_j} \in \mathbb{R}_{>0}$ is $E_j$'s maximum speed, and $\bm{u}_{E_j}$ belongs to $\mathbb{U}_E = \{ \bm{u} : [0, \infty) \rightarrow \mathbb{S}^2 \,|\, \bm{u} \textup{ is piecewise smooth}\}$. The initial position is $\bm{x}_{E_j}^0 = [x_{E_j}^0, y_{E_j}^0]^\top \in \mathbb{R}^2$. 

We denote the speed ratio between $P_i$ and $E_j$ by $\alpha_{ij}=v_{P_i}/v_{E_j}$ and consider faster pursuers, i.e., $\alpha_{ij} > 1$. The capture radius of pursuer $P_i$ is $r_i>0$. An evader is captured by a pursuer if the latter is pursuing the former and their Euclidean distance is less than or equal to the pursuer's capture radius. {\revise This is slightly different from the pioneering works \citep{AWM:71,AWM:74}  by Merz, where the capture occurs when their Euclidean distance is less than the capture radius for the convenience of determining the \emph{usable part} \citep{RI:65} of the terminal surface. However, since the approach we propose below is based on the forward analysis with no need to work retrogressively from the terminal surface, we here consider a closed capture condition such that the case when the distance is equal to the capture radius is also included.}  

The plane $\mathbb{R}^2$ is split by a closed convex curve $\targetline$ called the \emph{target curve}, into a \emph{goal region} $\goal$ and a \emph{play region} $\play$ as illustrated in Fig. \ref{fig:game-illustration}, and formally, they are respectively defined by 
\begin{equation*}
    \begin{aligned}
        \goal&=\{\bm{y}\in\mathbb{R}^2\,|\,g(\bm{y})\le0\}, \\
        \targetline&=\{\bm{y}\in\mathbb{R}^2\,|\,g(\bm{y})=0\}, \\
        \play&=\{\bm{y}\in\mathbb{R}^2\,|\,g(\bm{y})>0\},
    \end{aligned}
\end{equation*}
where $g: \mathbb{R}^2 \to \mathbb{R}$ is a twice differentiable function, i.e., $g_{\bm{y}} \triangleq  \od{g}{\bm{y}}$ and $g_{\bm{y}\bm{y}} \triangleq  \od[2]{g}{\bm{y}}$ exist, such that $\goal$ is nonempty, compact and convex. The evasion team $\eteam$ aims to send as many evaders initially in the play region as possible into the goal region before being captured by the pursuit team $\pteam$ who guards the goal region.

\subsection{Information structure}\label{subsec:information-structure}
{\revise The information available to each player plays an important role in determining game outcomes \citep{RI:65,RJE-NJK:72,IMM-AMB-CJT:05,PC:96}. Since this paper aims to propose strategies for the pursuit team $\pteam$, we will build the information structure from the pursuers' perspective. 
} We adopt the same information structure used in \cite{RY-RD-HL-WZ-ZS-YZ:23}. Under this information structure, the pursuit team makes decisions about its current control input with the information of all players’ current positions, plus the evasion team’s current control input (i.e., speeds and headings), {\revise which will be explicitly defined below}, while the evasion team is assumed to have only the access to
all players’ current positions. The maximum speeds of all players and the information about $\goal$, $\targetline$ and $\play$ are known by both teams.

{\revise As in \cite{RY-RD-HL-WZ-ZS-YZ:23}, we discuss how to practically access an evader's current control input (i.e., speed and heading under  \eqref{eq:evader_simple}).  Literature provides several methods to estimate the exogenous control signals of a moving object (such as an unmanned aerial vehicle), which can be used to estimate the evader's speed and heading in our problem~\citep{SB-TS:14,JM-NK-BB:20,RF-TS:18}, or assumes the evader's speed and heading are accessible~\citep{IE-PT-MP:15}. Advanced sensor systems, such as phased-array radars, can provide high-accuracy measurements of the velocity (i.e., speed and heading) of a moving object~\citep{soumekh1997phased,pihl2012phased}.}


\section{Enclosure Region Pursuit Winning and Coalition Reduction against One Evader}\label{sec:ERP-reduction}

It is hard to deal with multiple pursuers against multiple evaders directly due to complicated cooperation among team members and the lack of sensible matching rules between the players of different teams \citep{RY-XD-ZS-YZ-FB:19,MC-ZZ-CJT:17,DS-VK:20,DS-VK:18}. As an alternative, we split the whole game into many subgames, from which cooperative strategies and matching rules are extracted, thus comprising the team strategies. {\revise Such decomposition into subgames dates back to \cite{DL-JBC-GC-CK-MHC:05} where multi-player pursuit-evasion games were suboptimally solved based on the outcomes of all pairs of a pursuer and an evader, and was further generalised into a \emph{dynamic divide and conquer} approach \citep{VRM-PT:19}.}
The subgames considered in this section focus on multiple pursuers and a single evader.

We first introduce the following definitions and notations specialized for multiple pursuers. For any non-empty index set $c \in 2^{\{1,\dots,N_p\}}$, let $P_c=\{P_i \in \pteam \mid i \in c \}$ be an element of $2^{\pteam}$, and we refer to $P_c$ as a \emph{pursuit coalition} containing pursuer $P_i$ if $i \in c$. 
We denote by $X_c$ and $U_c$ the states (positions and headings) and control inputs of all pursuers in $P_c$, respectively, i.e., $X_c = \{(\bm{x}_{P_i}, \theta_{P_i})\}_{i \in c}$ and $U_c = \{ u_{P_i}\}_{i \in c}$. 

\subsection{Enclosure region pursuit winning}

Consider a subgame between a pursuit coalition $P_c$ and an evader $E_j$, in which $P_c$ wins the game if $E_j$ can never reach $\goal$ before being captured, while $E_j$ wins if it reaches $\goal$ prior to the capture. Our goal in this section is to determine, given the initial states, who wins the game. A complete solution to this qualitative problem involves solving an induced HJI equation numerically. However, the standard HJI method can only efficiently handle systems of up to five states \citep{MC-SLH-MSV-SB-CJT:18}, while this subgame has $3|c|+2$ states.
Note that $P_c$ wins the game in two possible ways: capture $E_j$ in $\play$ or infinitely delay $E_j$'s entry into $\play$. We instead establish a stronger winning condition that a dynamic region containing the evader consistently stays outside the goal region before the capture. This motivates our function-induced pursuit winning and strategies below. {\revise Actually, the core reasoning behind this winning condition was implicitly utilised in \cite{YL-EB:22,YL-EB:22-2,RY-XD-ZS-YZ-FB:19} when all players have the simple motion, where the dynamic region is Apollonius circle \citep{RI:65} or it generalisation.} We first introduce a class of functions. 

\begin{defi}\label{defi:PEF}{\rm (Pursuit enclosure function).}
For $P_i \in \pteam$ and $E_j \in \eteam$, let $X_{ij} = (\bm{x}_{P_i}, \theta_{P_i}, \bm{x}_{E_j})$ and $\mathcal{S}_{ij} = \{ X_{ij} \in \mathbb{R}^2 \times [0, 2\pi) \times \mathbb{R}^2 \mid \enVert[0]{\bm{x}_{P_i}-\bm{x}_{E_j}}_2 > r_i \}$. Then, a function $f:\mathbb{R}^2 \times \mathcal{S}_{ij}\to\mathbb{R}$ is a \emph{pursuit enclosure function (PEF)} if for each $X_{ij} \in \mathcal{S}_{ij}$, it satisfies the following conditions:
\begin{enumerate}
    \item\label{itm:PEF-defi-1} $\mathbb{E}(X_{ij};f)=\{\bm{x}\in\mathbb{R}^2\,|\,f(\bm{x},X_{ij})\ge0\}$ is compact and strictly convex. We call $\mathbb{E}(X_{ij};f)$ the \emph{enclosure region} of $P_i$ against $E_j$ via $f$;
    
    \item\label{itm:PEF-defi-2} 
    if $ f(\bm{x},X_{ij}) = 0$, then $f$ is differentiable in both $\bm{x}$ and $X_{ij}$;
    
    \item \label{itm:PEF-defi-3}$f(\bm{x}_{E_j},X_{ij})\ge0$.
\end{enumerate}
\end{defi}
 
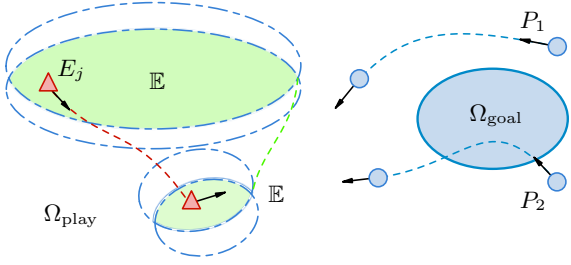
\begin{figure}
    \centering
\includegraphics[width=0.9\hsize]{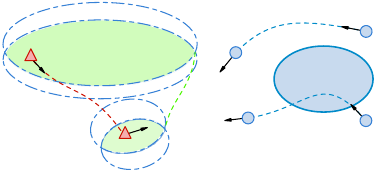}
    \put(-20,88){\footnotesize$P_1$}
    \put(-20,20){\footnotesize$P_2$}
    \put(-160,65){\footnotesize$\mathbb{E}$}
    \put(-195,70){\footnotesize$E_j$}
    \put(-115,22){\footnotesize$\mathbb{E}$}
    \put(-39,53){\footnotesize$\goal$}
    \put(-200,15){\footnotesize$\play$}
    \caption{If $P_1$ and $P_2$ can cooperatively ensure that the intersection $\mathbb{E}$ (green) of two enclosure regions containing $E_j$ never intersects with $\goal$, then the enclosure region pursuit (ERP) winning is achieved.}
    \label{fig:ERP-winning}
\end{figure}

The conditions (\ref{itm:PEF-defi-1}) and (\ref{itm:PEF-defi-3}) imply that the evader $E_j$ is contained in the enclosure region $\mathbb{E}(X_{ij};f)$ ({\revise Fig. \ref{fig:ERP-winning} illustrates four enclosure regions, each of which has a dotted boundary}).
We are now ready to define a new class of pursuit winning conditions based on the PEFs. Let $X_{cj}=(X_c,\bm{x}_{E_j})$ be the state of the system comprising of $P_c$ and $E_j$, and $\mathcal{S}_{cj}= \{ X_{cj} \mid \forall i \in c, \|\bm{x}_{P_i} - \bm{x}_{E_j}\|_2 > r_i \} $ be the set of states such that $E_j$ is not captured by any pursuer in $P_c$. 
 Unless otherwise stated we assume that $X_{cj} \in \mathcal{S}_{cj}$ hereinafter. {\revise In this subgame, a strategy of $P_i$ ($i \in c$) under the information structure in Section~\ref{subsec:information-structure} is a mapping $u_{P_i} : \mathcal{S}_{cj} \times \mathbb{S}^2 \to \mathbb{S}^1$, and a strategy of $E_j$ is a mapping $\bm{u}_{E_j} : \mathcal{S}_{cj} \to \mathbb{S}^2$. Note that the strategy $u_{P_i}$ requires the current control input of $E_j$ (i.e., speed and heading), but not the strategy of $E_j$.}
 
 Suppose that $P_c$ is endowed with a set $f_c$ of PEFs, where $f_c = \{f_i\}_{i \in c}$ and $f_i$ is a PEF for pursuer $P_i$ $(i \in c)$, and we omit the subscript $j$ for PEFs for simplicity. 
 Then, the intersection of the enclosure regions of $P_i$ against $E_j$ via $f_i$ for all $i \in c$, is
 \begin{equation*}
\begin{aligned}
   \mathbb{E}(X_{cj}; f_c)  &  \triangleq \cap_{i \in c}\mathbb{E}(X_{ij};f_i) \\
   & = \{\bm{x}\in\mathbb{R}^2 \,|\, f_i(\bm{x},X_{ij}) \ge 0, i \in c\}.
\end{aligned}
 \end{equation*}
{\revise The green region in Fig. \ref{fig:ERP-winning} is the intersection of two enclosure regions of $P_1$ and $P_2$ against $E_j$.} By \defiref{defi:PEF}, $\mathbb{E}(X_{cj}; f_c)$ is nonempty because $\bm{x}_{E_j} \in \mathbb{E}(X_{cj}; f_c)$, and it is compact and strictly convex. Letting
\[\varrho(X_{cj}; f_c) \triangleq d( \mathbb{E}(X_{cj}; f_c), \goal )\]
be the distance between the region $\mathbb{E}(X_{cj}; f_c)$ and the goal region $\goal$, we  next introduce \emph{safe distance} for $P_c$ against $E_j$.

\begin{defi}\label{defi:safe-distance}{\rm (Safe distance).} Consider a pursuit coalition $P_c$ and an evader $E_j$. Given a set $f_c$ of PEFs, the \emph{safe distance} of a state $X_{cj} \in \mathcal{S}_{cj}$ under $f_c$ is $\varrho (X_{cj}; f_c)$.
\end{defi}
Since $\bm{x}_{E_j} \in \mathbb{E}(X_{cj}; f_c)$, then $d(\bm{x}_{E_j}, \goal) \ge \varrho (X_{cj}; f_c)$ for all $X_{cj} \in \mathcal{S}_{cj}$. This implies that if the safe distance is positive, i.e., $\varrho (X_{cj}; f_c) > 0$, then $E_j$ resides outside of $\goal$. Let $X_{cj}^{t}$ be the system state at $t \ge 0$ starting from $X_{cj}$ under control inputs $U_c$ and $\bm{u}_{E_j}$.
We next introduce a new pursuit winning strategy using the safe distance.

\begin{defi}{\rm(ERP winning state and strategy).}\label{defi:ERP-winning}
Given a set $f_c$ of PEFs, a state $X_{cj} \in \mathcal{S}_{cj}$ is an enclosure region pursuit (ERP) winning state, if there exists a pursuit strategy $U_c$ for $P_c$ such that the safe distance is positive from $X_{cj}$ for all $t \ge 0$, i.e., $\varrho (X_{cj}^t; f_c)  > 0$ for all $t \ge 0$, regardless of $\bm{u}_{E_j}$. Such a strategy $U_c$ is called an ERP winning strategy. 
\end{defi}

By \defiref{defi:ERP-winning}, if a state $X_{cj}$ is an ERP winning state, then the pursuit coalition can ensure that, from this state, the evader can never enter $\goal$ prior to the capture, by using the ERP winning strategy, as depicted in Fig. \ref{fig:ERP-winning}. Since $\mathbb{E}(X_{cj}; f_c)$ is strictly convex and $\goal$ is convex, computing the safe distance $\varrho (X_{cj}; f_c)$ involves solving a convex optimization problem.

\begin{defi}\label{defi:closest-point-mapping}{\rm (Computing the safe distance).}
Given a set $f_c$ of PEFs and a state $X_{cj} \in \mathcal{S}_{cj}$, let $(\bm{x}_I, \bm{x}_G)$ be the solution of the convex optimization problem $\mathcal{P}(X_{cj})$:
\begin{equation}\label{eq:convex-pbm-IG}
\begin{aligned}
    & \underset{(\bm{x},\bm{y})\in\mathbb{R}^2\times\mathbb{R}^2}{\textup{minimize}}
	&& d (\bm{x},\bm{y})\\
	&\textup{ subject to}&& f_i(\bm{x},X_{ij})\ge 0,\; g(\bm{y})\le 0, \; \forall i\in c.
\end{aligned}
\end{equation}
Then, the safe distance is $\varrho (X_{cj}; f_c) = d (\bm{x}_I, \bm{x}_G)$.
\end{defi}

{\revise A recent relevant work by Lee and Bakolas \citep{YL-EB:22}, which focuses on  simple-motion players in $\mathbb{R}^n$ with point capture, requires computing the distance between two convex sets and has proposed an interesting alternating projection algorithm to solve an induced convex optimization problem similar to \eqref{eq:convex-pbm-IG}. This alternating projection algorithm can also be used to solve \eqref{eq:convex-pbm-IG} with a slight modification.}

\subsection{Coalition reduction}
The convex optimization problem $\mathcal{P}(X_{cj})$ in \eqref{eq:convex-pbm-IG} is used to compute the safe distance which plays a vital role in the following analysis. Noting this, we next prove that given $X_{cj}$, there are at most two PEFs in $f_c$ which are necessary to solve $\mathcal{P}(X_{cj})$, greatly simplifying the computation. This also implies that at most two pursuers in $P_c$ work when computing $\bm{x}_I$. {\revise Similar coalition reduction for simple-motion players can be found in \cite{AVM-EG-DC-MS-SCS:20,RY-ZS-YZ:20-1,RY-XD-ZS-YZ-FB:19}.}

Instead of concentrating on the specific convex problem \eqref{eq:convex-pbm-IG}, we consider support constraints (defined later) for more general convex optimization problems. Consider the optimization problem
\begin{equation}\label{eq:convex-pbm-support}
    \mathcal{P} : \; \underset{\bm{x}\in \mathbb{R}^n }{\textup{minimize}} \; \phi(\bm{x}) \quad \textup{subject to :} \; \bm{x} \in \bigcap_{i \in \{1,\dots,m\}} \mathcal{X}_i
\end{equation}
where $\phi(\bm{x})$ is a real-valued function in $\bm{x}$ and $\mathcal{X}_i$ ($i=1,\dots,m$) are closed convex sets. Then, define the programs $\mathcal{P}_k$, $k=1,\dots,m$ obtained from $\mathcal{P}$ by removing the $k$th constraint
\begin{equation*}
    \mathcal{P}_k : \; \underset{\bm{x}\in \mathbb{R}^n }{\textup{minimize}} \; \phi(\bm{x}) \quad \textup{subject to :} \; \bm{x} \in \bigcap_{i \in \{1,\dots,m\} \setminus k} \mathcal{X}_i.
\end{equation*}
We assume program $\mathcal{P}$ and the programs $\mathcal{P}_k$ admit an optimal solution, say $\bm{x}^*$ and $\bm{x}^*_k$, respectively, and let $J^*=\phi(\bm{x}^*)$ and $J^*_k=\phi(\bm{x}^*_k)$. 

\begin{defi}{\rm (Support constraint).}
The $k$th constraint $\mathcal{X}_k$ is a \emph{support constraint} for $\mathcal{P}$ if $J_k^*<J^*$.
\end{defi}

We first recall a well-known result due to Calafiore and Campi; see \cite{GCC-MCC:06}.

\begin{lema}\label{lema:calafiore}{\rm (Calafiore and Campi).}
If $\phi(\bm{x})=\bm{c}^{\top}\bm{x}$ and program $\mathcal{P}$ and the programs $\mathcal{P}_k$ admit a unique optimal solution, then the number of support constraints for problem $\mathcal{P}$ is at most $n$.
\end{lema}

By \lemaref{lema:calafiore}, the convex program $\mathcal{P}(X_{cj})$ in \eqref{eq:convex-pbm-IG} can be solved through a convex program with fewer constraints.

\begin{thom}\label{lema:constraint-reduction}{\rm (Constraint reduction).} Let $J^*_{cj}$ be the optimal value of the convex program $\mathcal{P}(X_{cj})$. Then, if $J^*_{cj} > 0$, there exists a subcoalition $\bar{c}$ of $c$ such that $|\bar{c}| \leq 2$ and $J_{cj}^* = J^*_{\bar{c}j}$, where $J^*_{\bar{c}j}$ is the optimal value of the convex optimization problem $\mathcal{P}(X_{\bar{c}j})$:
\begin{equation}\label{eq:convex-pbm-IG-reduction}
\begin{aligned}
 & \underset{(\bm{x},\bm{y})\in\mathbb{R}^2\times\mathbb{R}^2}{\textup{minimize}}	&& d (\bm{x},\bm{y})\\
	&\textup{ subject to}&& f_i(\bm{x},X_{ij})\ge 0,\; g(\bm{y})\le 0, \; \forall i\in \bar{c}.
\end{aligned}
\end{equation}

Proof. \rm {\revise We prove the theorem by formulating the convex program $\mathcal{P}(X_{cj})$ in \eqref{eq:convex-pbm-IG} as a special case of \eqref{eq:convex-pbm-support} and then using \lemaref{lema:calafiore} to reduce the constraints, which leads to~\eqref{eq:convex-pbm-IG-reduction}.} Since $J^*_{cj} > 0$, then $\varrho (X_{cj}; f_c)  = d (\bm{x}_I, \bm{x}_G)  > 0$ and therefore $d( \mathbb{E}(X_{cj}; f_c), \goal ) > 0$, where $(\bm{x}_I, \bm{x}_G)$ is the optimal solution to $\mathcal{P}(X_{cj})$. Since $\mathbb{E}(X_{cj}; f_c)$ is closed and strictly convex and $\goal$ is closed and convex, then the optimal solution $(\bm{x}_I, \bm{x}_G)$ is unique.

Then, $\mathcal{P}(X_{cj})$ in \eqref{eq:convex-pbm-IG}  is equivalent to the following problem $\overline{\mathcal{P}}$:
\begin{equation}\label{eq:new-P-cj}
\begin{aligned}
& \underset{(\bm{x}, \gamma) \in \mathbb{R}^2 \times \mathbb{R}}{\textup{minimize}}
	&& \gamma \\
	& \textup{subject to}&& f_i(\bm{x},X_{ij})\ge 0, \; \forall i\in c,\\
    & & & d (\bm{x}, \bm{x}_G) - \gamma \leq 0 .
\end{aligned}
\end{equation}
In \eqref{eq:new-P-cj}, $(\bm{x}, \gamma)$ is three-dimensional, and there are $|c| + 1$ convex constraints as the convexity of $d (\bm{x}, \bm{x}_G) - \gamma \leq 0$ is straightforward. In order to apply \lemaref{lema:calafiore}, we next show the uniqueness of the solution. We  only consider the case $|c| \ge 3$, as we can take $\bar{c} = c$ when $|c|\leq 2$.

Since $\mathcal{P}(X_{cj})$ has the unique optimal solution $(\bm{x}_I, \bm{x}_G)$, then $\overline{\mathcal{P}}$ has a unique optimal solution $(\bm{x}_I, \gamma^*)$, where $\gamma^* = d (\bm{x}_I, \bm{x}_G)$. For $i \in c$, we let $\overline{\mathcal{P}}_i$ be the convex program obtained from $\overline{\mathcal{P}}$ by removing the constraint $f_i(\bm{x},X_{ij})\ge 0$. Then, following the same argument to $\overline{\mathcal{P}}$, we have that $\overline{\mathcal{P}}_i$ admits a unique optimal solution.
We do not need to remove the constraint $d (\bm{x}, \bm{x}_G) - \gamma \leq 0$ as it is a support constraint for $\overline{\mathcal{P}}$.

By \lemaref{lema:calafiore}, we conclude that $\overline{\mathcal{P}}$ has at most 3 support constraints. Since $d (\bm{x}, \bm{x}_G) - \gamma \leq 0$ is a support constraint,
then there are at most two support constraints among $f_i(\bm{x},X_{ij})\ge 0$ for $i \in c$. \qed 
\end{thom}

\section{ERP Winning State Generation}\label{sec:sink-winning-subspace}

If a set of states are known to be ERP winning, establishing the connection between this set and other states in $\mathcal{S}_{cj}$ can help find more ERP winning states. This section proposes a \emph{steer-to-ERP} approach to verify whether a state is ERP winning based on the known ERP winning states, via steering the system to a known ERP winning state.

The following notations will be required. For a PEF $f_i$, we define $f_{i, \bm{x}}(\bm{x}, X_{ij}) = \partial f (\bm{x}, X_{ij})/ \partial \bm{x}$, $f_{i, P}(\bm{x}, X_{ij}) = \partial f (\bm{x}, X_{ij})/ \partial \bm{x}_{P_i}$, $f_{i, \theta}(\bm{x}, X_{ij}) = \partial f (\bm{x}, X_{ij}) / \partial \theta_{P_i} $, and $f_{i, E}(\bm{x}, X_{ij}) = \partial f (\bm{x}, X_{ij}) / \partial \bm{x}_{E_j} $. For two distinct points $\bm{x} \in \mathbb{R}^2 $ and $\bm{y} \in \mathbb{R}^2$, let $d_{\bm{x}} (\bm{x}, \bm{y}) = \partial d (\bm{x}, \bm{y}) / \partial \bm{x}$ and $d_{\bm{y}} (\bm{x}, \bm{y}) = \partial d (\bm{x}, \bm{y}) / \partial \bm{y}$. Let $\bm{e}_{\theta_{P_i}} = [\cos \theta_{P_i}, \sin \theta_{P_i}]^\top$.

\begin{thom}\label{thom:nece-suff-no-closing-nv1}{\rm (Steer-to-ERP approach).} 
Given a set $f_c$ of PEFs and a state $X_{cj} \in \mathcal{S}_{cj}$, if there exists a pursuit strategy $U_c^1$ of $P_c$ and a time horizon $T \in \mathbb{R}_{\ge 0}$ such that $X_{cj}^{t^{\star}}$ is an ERP winning state for some $t^{\star} \in [0, T]$ regardless of $\bm{u}_{E_j}$, and $\varrho(X_{cj}; f_c) > \max \{-V T, 0\} $, where $V$ is the optimal value of problem
\begin{equation}\label{eq:optimization-pbm-set-nv1}
    \begin{aligned}
       & \textup{minimize} & &  \sum\nolimits_{i \in c}  v_{P_i} \lambda_i \big ( f_{i,P}^\top \bm{e}_{\theta_{P_i}} + \frac{  |f_{i,\theta}| }{\kappa_i } + \frac{\| f_{i,E} \|_2}{\alpha_{ij}} \big) \\
       & \textup{variables}&& \bm{x}_I \in \mathbb{R}^2 ,\bm{x}_G \in \mathbb{R}^2, X'_{cj} \in \mathcal{S}_{cj}, \lambda_i, \lambda_g, i \in c \\
       &\textup{subject to} & & \bm{0}= \lambda_g g_{\bm{y}}(\bm{x}_G)  + \sum\nolimits_{i \in c} \lambda_i f_{i, \bm{x}} (\bm{x}_I, X'_{ij} ) \\
       & & & \bm{0}= d_{\bm{y}} ( \bm{x}_I, \bm{x}_G) + \lambda_g g_{\bm{y}}(\bm{x}_G) \\
       & & & f_i ( \bm{x}_I, X'_{ij}) \ge 0, \lambda_i \leq 0, \lambda_i f_i ( \bm{x}_I, X'_{ij}) = 0 \\
       & & &  i \in c, g(\bm{x}_G) = 0,  \lambda_g \ge 0
    \end{aligned}
\end{equation}
then $X_{cj}$ is an ERP winning state. Moreover, if $U_c^2$ is an ERP winning strategy for $X_{cj}^{t^{\star}}$, then taking $U_c^1$ for $[0, t^{\star}]$ and $U_c^2$ for $(t^{\star}, \infty)$ is an ERP winning strategy for $X_{cj}$.

Proof. \rm
{\revise Since $X_{cj}$ reaches an ERP winning state $X_{cj}^{t^{\star}}$ for some $t^{\star} \in [0, T]$ under $U_c^1$ regardless of $\bm{u}_{E_j}$, then $X_{cj}$ is an ERP winning sate if the positive safe distance is kept before $X_{cj}^{t^{\star}}$ is reached. To that end, we first compute the minimum speed of the intersection of enclosure regions moving away from $\goal$ and then obtain the minimum (possibly negative) increment of the safe distance during the time horizon $T$. Thus, $X_{cj}$ is an ERP winning sate if the initial safe distance $\varrho(X_{cj}; f_c)$ can compensate this minimum increment.}

Consider a state $X'_{cj} \in \mathcal{S}_{cj}$ such that $\varrho(X'_{cj}; f_c) > 0$, that is, $d (\mathbb{E}(X'_{cj}; f_c), \goal) > 0$. Since $\mathbb{E}(X'_{cj}; f_c)$ is strictly convex and $\goal$ is convex, there exists a unique solution $(\bm{x}_I, \bm{x}_G) \in \mathbb{E}(X'_{cj}; f_c) \times \goal$ to $\mathcal{P}(X'_{cj})$ in \eqref{eq:convex-pbm-IG}, {\revise and we also have $d(\bm{x}_I, \bm{x}_G) > 0$.}

By \emph{Karush-Kuhn-Tucker (KKT) conditions} {\revise for \eqref{eq:convex-pbm-IG}}, the solution $(\bm{x}_I, \bm{x}_G)$ satisfies
\begin{subequations}\label{eq:kkt-condition-nv1}
\begin{align}
	&\bm{0}= d_{\bm{x}} (\bm{x}_I, \bm{x}_G) + \sum\nolimits_{i \in c} \lambda_i f_{i, \bm{x}} (\bm{x}_I, X'_{ij}) \label{itm:winning-KKT-1}\\
	&\bm{0}= d_{\bm{y}} (\bm{x}_I, \bm{x}_G) + \lambda_g g_{\bm{y}}(\bm{x}_G) \label{itm:winning-KKT-2}\\
	& f_i (\bm{x}_I, X'_{ij}) \ge 0, \ \lambda_i \leq 0, \ \lambda_i f_i(\bm{x}_I, X'_{ij}) = 0,  \ i \in c \label{itm:winning-KKT-3}\\
	& g(\bm{x}_G)\le 0, \ \lambda_g g(\bm{x}_G) = 0, \ \lambda_g \ge 0 \label{itm:winning-KKT-4}
\end{align}
\end{subequations}
where $X'_{ij}$ is a part of $X'_{cj}$ corresponding to $P_i$, and $\lambda_i$ and $ \lambda_g$ are the Lagrange multipliers. The complementary 
 slackness condition \eqref{itm:winning-KKT-3} implies that the index set $c$ can be classified into two disjoint index sets $c^{=0}$ and $c^{>0}$ ($c^{>0}$ may be empty) where
\begin{equation}\label{eq:slack-condition}
\begin{cases}
f_i(\bm{x}_I, X'_{ij}) = 0, \ \lambda_i \leq 0, \quad \textup{if } i \in c^{=0}, \\
f_i(\bm{x}_I, X'_{ij}) > 0, \ \lambda_i = 0, \quad \textup{if } i \in c^{>0}.
\end{cases}
\end{equation} 
Then the region $\mathbb{E}(X'_{cj}; f_c)$ moves away from $\goal$ with the speed $\od{}{t}{\varrho} (X'_{cj}; f_c)$ that equals
\begin{equation}\label{eq:varrho-to-f-nv1}
\begin{aligned}
	 & \dod{}{t}d ( \bm{x}_I, \bm{x}_G ) =  d_{\bm{x}}^{\top}( \bm{x}_I, \bm{x}_G ) \dot{\bm{x}}_I + d_{\bm{y}}^{\top} ( \bm{x}_I, \bm{x}_G ) \dot{\bm{x}}_G \\
	& = - \sum\nolimits_{i \in c} \lambda_i f^{\top}_{i, \bm{x}} (\bm{x}_I, X'_{ij} ) \dot{\bm{x}}_I - \lambda_g g_{\bm{y}}^{\top} (\bm{x}_G ) \dot{\bm{x}}_G \\
	& = - \sum\nolimits_{i \in c} \lambda_i f^{\top}_{i, \bm{x}} (\bm{x}_I, X'_{ij} ) \dot{\bm{x}}_I,
\end{aligned}
\end{equation}
where 
the second equality is due to \eqref{eq:kkt-condition-nv1}, and the third equality follows noting that $\bm{x}_G$ is always at the boundary of $\goal$, i.e., $g(\bm{x}_G) \equiv 0$ and thus $g_{\bm{y}}^{\top} (\bm{x}_G) \dot{\bm{x}}_G = 0$. For any $i \in c^{=0}$, since $\bm{x}_I$ is always at the boundary of $\mathbb{E}(X'_{ij}; f_i)$, then $f_i(\bm{x}_I, X'_{ij}) \equiv 0$,  and according to the condition \ref{itm:PEF-defi-2} in \defiref{defi:PEF}, $f_i$ is differentiable in $\bm{x}$, $\bm{x}_{P_i}$, $\theta_{P_i}$ and $\bm{x}_{E_j}$. Then, we have $\od{}{t}f_i(\bm{x}_I, X'_{ij})=0$, implying that
\begin{equation}\label{eq:f-to-control-nv1}
\begin{aligned}
& f_{i,\bm{x}}^\top (\bm{x}_I, X'_{ij}) \dot{\bm{x}}_I  = - f_{i,P}^\top \dot{\bm{x}}_{P_i} - f_{i,\theta} \dot{\theta}_{P_i} - f_{i,E}^\top  \dot{\bm{x}}_{E_j}\\
& = - v_{P_i} f_{i,P}^\top \bm{e}_{\theta_{P_i}} - f_{i,\theta} v_{P_i} u_{P_i} / \kappa_i - v_{E_j} f_{i,E}^\top  \bm{u}_{E_j},
\end{aligned}
\end{equation}
where the second equality follows from \eqref{eq:pursuer_car} and \eqref{eq:evader_simple}. 

Then, the minimum speed of $\mathbb{E}(X'_{cj}; f_c)$ moving away from $\goal$, denoted by $V(X'_{cj})$, is equal to
\begin{equation*} 
\begin{aligned}
& \min_{U_c \in \mathbb{U}_P^c} \min_{\bm{u}_{E_j} \in \mathbb{U}_E}  \dod{}{t}{\varrho}(X'_{cj}; f_c) \\ 
     &  = \sum\nolimits_{i \in c}  \lambda_i ( v_{P_i} f_{i,P}^\top \bm{e}_{\theta_{P_i}} + |f_{i,\theta}| v_{P_i} / \kappa_i + v_{E_j} \| f_{i,E} \|_2),
\end{aligned}    
\end{equation*}
where \eqref{eq:varrho-to-f-nv1}, \eqref{eq:f-to-control-nv1} and $\lambda_i \leq 0$ are used. The constraint \eqref{eq:kkt-condition-nv1} is written as \eqref{eq:optimization-pbm-set-nv1} using $d_{\bm{x}} (\bm{x}_I, \bm{x}_G) = - d_{\bm{y}} (\bm{x}_I, \bm{x}_G)$. 
Thus, the minimum speed of the region $\mathbb{E}(X'_{cj}; f_c)$ moving away from $\goal$ over $\mathcal{S}_{cj}$ is $V = \min_{X'_{cj} \in \mathcal{S}_{cj}} V(X'_{cj})$. Moreover, by assumption $P_c$ can steer the system state $X_{cj}$ to an ERP winning state $X_{cj}^{t^{\star}}$ within the time period $T$ by using the strategy $U_c^1$.
Thus, if the current system state $X_{cj} \in \mathcal{S}_{cj}$ is such that $\varrho(X_{cj}; f_c) > \max \{-VT, 0\} $, then the positive safe distance is kept before $X_{cj}^{t^{\star}}$ is reached. Thus, by \defiref{defi:ERP-winning}, $X_{cj}$ is an ERP winning state. If $U_c^2$ is an ERP winning strategy for $X_{cj}^{t^{\star}}$, then the strategy pair $(U_c^1, U_c^2)$ can form an ERP winning strategy for $X_{cj}$ as described. \qed

\begin{rek}
\label{rek:steer-to-ERP-speed}
By the proof of \thomref{thom:nece-suff-no-closing-nv1}, the optimal value $V$ to the problem \eqref{eq:optimization-pbm-set-nv1} is the minimum speed of the region $\mathbb{E}(X'_{cj}; f_c)$ moving away from $\goal$ for all $X'_{cj} \in \mathcal{S}_{cj}$. Moreover, the nonlinear program \eqref{eq:optimization-pbm-set-nv1} can be largely simplified if the derivatives of the PEFs are easy to compute.  
\end{rek}
\end{thom}

\section{Positional Pursuit Enclosure Function}\label{sec:positional-PEF}

This section introduces a class of PEFs based on players' positions which extend the potential function in \cite{RY-XD-ZS-YZ-FB:19} to the plane. We present an initial set of the induced ERP winning states and then generate more based on them using the steer-to-ERP approach. Since the coalition reduction in \thomref{lema:constraint-reduction} shows that at most two pursuers are needed to ensure an ERP winning against an evader, we present the parameters, states and strategies that can ensure the ERP winning for the cases of one pursuer and two pursuers, respectively.

\subsection{Positional PEFs}

The following lemma identifies the positional PEF.
\begin{lema}{\rm (Positional PEF).} For $P_i \in \pteam$ and $E_j \in \eteam$, 
the function 
$f_i^{\textup{ps}} : \mathbb{R}^2 \times \mathcal{S}_{ij} \to \mathbb{R}$ defined by
\begin{equation}\label{eq:positional-PEF}
\begin{aligned}
    f_i^{\textup{ps}}(\bm{x},  X_{ij}) =\enVert[0]{\bm{x}-\bm{x}_{P_i}}_2 - \alpha_{ij} \enVert[0]{\bm{x}-\bm{x}_{E_j}}_2 - r_i
\end{aligned}
\end{equation}
is a PEF.
We call $f_i^{\textup{ps}}$ the positional PEF.

Proof. \rm
We prove that $f_i^{\textup{ps}}$ is a PEF by verifying the conditions (\ref{itm:PEF-defi-1})-(\ref{itm:PEF-defi-3}) in \defiref{defi:PEF}. Regarding the condition (\ref{itm:PEF-defi-2}), 
if $\bm{x} = \bm{x}_{P_i}$, then $f_i^{\textup{ps}} =  - \alpha_{ij} \enVert[0]{\bm{x}_{P_i}-\bm{x}_{E_j}}_2 - r_i < 0$.
If $\bm{x} = \bm{x}_{E_j}$, then $f_i^{\textup{ps}}  = \enVert[0]{\bm{x}_{E_j}-\bm{x}_{P_i}}_2 - r_i > 0$ noting $X_{ij} \in \mathcal{S}_{ij}$. Therefore, if $f_i^{\textup{ps}} = 0$, then $\bm{x} \neq \bm{x}_{P_i}$ and $\bm{x} \neq \bm{x}_{E_j}$, and thus $f_i^{\textup{ps}}$ is differentiable in $\bm{x}$, $\bm{x}_{P_i}$, $\theta_{P_i}$ and $\bm{x}_{E_j}$. Regarding the condition \eqref{itm:PEF-defi-3}, we have $f_i^{\textup{ps}}(\bm{x}_{E_j}, X_{ij}) = \enVert[0]{\bm{x}_{E_j}-\bm{x}_{P_i}}_2 - r_i > 0$.

Regarding the condition \eqref{itm:PEF-defi-1}, we build a polar coordinate system with $\bm{x}_{E_j}$ as the origin, and let $\bm{x} = \bm{x}_{E_j} + \rho \bm{e}$, where $\rho \in \mathbb{R}_{> 0}$ and $\bm{e} \in \partial \mathbb{S}^2$. We parameterize $\bm{e}$ by $\bm{e} = (\cos (\psi + \psi_0), \sin (\psi + \psi_0))$, where $\psi \in [0, 2 \pi)$ is the rotation with respect to positive $x$-axis, and $\psi_0 \in [0, 2 \pi)$ is the initial rotation. Then, the boundary of $\mathbb{E}(X_{ij};f_i^{\textup{ps}})$, i.e., $f_i^{\textup{ps}}(\bm{x}, X_{ij}) = 0$, in this polar coordinate becomes $\| \bm{x}_{E_j} + \rho \bm{e} - \bm{x}_{P_i} \|_2 - \alpha_{ij} \rho - r_i = 0$. Thus we have $\rho = \frac{1}{\alpha_{ij}^2 - 1} (h_1(\psi) + h_2(\psi))$, where $h_1(\psi)$ and $h_2(\psi)$ are
\begin{equation}\label{eq:h1-h2}
\begin{aligned}
   h_1(\psi) & = (\bm{x}_{E_j} - \bm{x}_{P_i})^{\top} \bm{e} - \alpha_{ij} r_i \\
   h_2(\psi) & = \sqrt{ h_1^2(\psi) + (\alpha_{ij}^2 - 1)(\|\bm{x}_{E_j} - \bm{x}_{P_i}\|_2^2 - r_i^2)}.
\end{aligned}
\end{equation}
In deriving \eqref{eq:h1-h2}, $\alpha_{ij} > 1$ and $\|\bm{x}_{E_j} - \bm{x}_{P_i}\|_2 > r_i$ are used, which also implies that $h_2 > 0$ and $h_2$ is real as opposed to complex. Thus, given $X_{ij}$, $\rho$ is bounded and $\rho > 0$, and thus $\mathbb{E}(X_{ij};f_i^{\textup{ps}})$ is bounded. Since the boundary is contained in $\mathbb{E}(X_{ij};f_i^{\textup{ps}})$, then it is compact. As for the strict convexity, following the same argument in the proof of \cite[Lemma 3.1]{RY-XD-ZS-YZ-FB:19}, we have $ \rho^2 + 2 (\od{\rho}{\psi})^2 - \rho \od[2]{\rho}{\psi} > 0$ for all $\psi$. By \cite[Lemma 2.1]{RY-XD-ZS-YZ-FB:19}, $\mathbb{E}(X_{ij};f_i^{\textup{ps}})$ is strictly convex. \qed

\end{lema}

{\revise Given $X_{ij} \in \mathcal{S}_{ij}$, the locus of $f_i^{\textup{ps}}(\bm{x},  X_{ij}) = 0$ is a Cartesian oval, also called Apollonius oval \citep{PW-MP-KP:19}.} Suppose that each pursuer in $P_c$ adopts the positional PEF and let $f_c^{\textup{ps}} = \{ f_i^{\textup{ps}} \}_{i \in c}$. We consider a set of states: 
\begin{equation}\label{eq:position-initial-ERP}
 \mathcal{X}_{cj}^{\textup{ps}} = \{ X_{cj} \in \mathcal{S}_{cj} | \varrho(X_{cj}; f_c^{\textup{ps}}) > 0, \theta_{P_i} = \sigma_i(X_{cj}), i \in c \}   
\end{equation}
where $\sigma_i : \mathcal{S}_{cj} \to [0, 2\pi) $ is a heading function such that for each $X_{cj} \in \mathcal{S}_{cj}$, $[\cos\sigma_i(X_{cj}), \sin \sigma_i(X_{cj})]^\top = \frac{\bm{x}_I - \bm{x}_{P_i}}{\| \bm{x}_I - \bm{x}_{P_i} \|_2}$, where $(\bm{x}_I, \bm{x}_G)$ is the optimal solution of the convex problem $\mathcal{P}^{\textup{ps}}(X_{cj})$:
\begin{equation}\label{eq:convex-pbm-AC-PEF}
\begin{aligned}
    & \underset{(\bm{x},\bm{y})\in\mathbb{R}^2\times\mathbb{R}^2}{\textup{minimize}}
	&& d(\bm{x},\bm{y})\\
	&\textup{ subject to}&& f_i^{\textup{ps}}(\bm{x}, X_{ij}) \ge 0, \ g(\bm{y})\le 0, \ i \in c.
\end{aligned}
\end{equation}
Next, we present the conditions on the parameters such that $\mathcal{X}_{cj}^{\textup{ps}}$ is a set of ERP winning states and generate more from them via the steer-to-ERP approach, for both one-pursuer and two-pursuer cases.

\subsection{ERP winning conditions for one pursuer}

We introduce several notations first. We let $d_{opt_1opt_2} = \| \bm{x}_{opt_1} - \bm{x}_{opt_2} \|_2$ (interchangeable with $d(\bm{x}_{opt_1}, \bm{x}_{opt_2})$) and $\bm{e}_{opt_1opt_2} = (\bm{x}_{opt_1} - \bm{x}_{opt_2} ) / d_{opt_1opt_2}$ (if $d_{opt_1opt_2} > 0$) for $opt_1, opt_2 \in \{I, G, P_i \in \pteam, E_j \in \eteam \}$. For the goal region $\goal$, let $\bm{e}_{IG} = g_{\bm{y}}(\bm{x}_G) / \| g_{\bm{y}}(\bm{x}_G) \|_2$ and $H(\bm{x}_G)$ be the unit gradient and the Hessian matrix of $g$ at $\bm{x}_G$, respectively.

We first consider the case of one pursuer and one evader. Based on $\mathcal{X}_{cj}^{\textup{ps}}$, we present the conditions on parameters and states that can ensure the ERP winning via the steer-to-ERP approach, and give the corresponding ERP winning strategies.

\begin{thom}\label{thom:sink-winning-AC-1v1}{\rm (ERP winning parameters, state and strategy).}
Consider a one-pursuer pursuit coalition $P_c = \{ P_i \} $ against an evader $E_j$.
If the following conditions hold:
\begin{enumerate}
    \item the parameters satisfy
\begin{equation}\label{eq:parameters-1v1}
    \alpha_{ij} > 3, 
    \quad r_i / \kappa_i > \textup{CM}_1(\alpha_{ij})
\end{equation}
where
$\textup{CM}_1(\alpha_{ij})$ is a bound of the ratio between the capture radius and the minimum turning radius:
\begin{equation}
    \textup{CM}_1 (\alpha_{ij}) = 1 + \frac{ 3 \alpha_{ij}^2 + 4 \alpha_{ij} - 3}{(\alpha_{ij} - 1)^2 (\alpha_{ij} -3)};
\end{equation}

\item the safe distance of the state $X_{ij} \in \mathcal{S}_{ij}$ satisfies
\begin{equation}\label{eq:configuration-1v1}
    \varrho(X_{ij}; f_i^{\textup{ps}} ) >  \frac{ 2 \pi r_i / (\alpha_{ij} - 1) }{ r_i / \kappa_i - \textup{CM}_1(\alpha_{ij})}
\end{equation}

\end{enumerate}
then $X_{ij}$ is an ERP winning state and the feedback pursuit strategy 
\begin{equation}\label{eq:sink-strategy-1v1}
u_{P_i} = 
\begin{cases}
- \frac{\kappa_i}{v_{P_i}} \frac{\bm{e}_{IP_i}^{\circ \top} \dot{\bm{x}}_I}{d_{IP_i} }, \ \textup{if } \theta_{P_i} = \sigma_i(X_{ij})\\
\textup{sgn} (\sin (\sigma_i(X_{ij}) - \theta_{P_i})), \ \textup{otherwise } 
\end{cases}
\end{equation}
for $X_{ij} \in \mathcal{S}_{ij}$, is an ERP winning strategy with $(\bm{x}_I, \bm{x}_G)$ computed by the convex optimization problem \eqref{eq:convex-pbm-AC-PEF} and 
\begin{equation*}
    \dot{\bm{x}}_I = \frac{ (\alpha_{ij} \bm{a}_1^{\circ} \bm{e}^\top_{IE_j} + A)  \dot{\bm{x}}_{E_j} 
- v_{P_i} \bm{a}_1^{\circ} }{
d_f \bm{a}_1^\top  \bm{e}_{IG}^{\circ}
}
\end{equation*}
where $d_{f} = \| \bm{e}_{IP_i} - \alpha_{ij} \bm{e}_{IE_j} \|_2$ and
\begin{equation}\label{eq:b1-B-1v1}
\begin{aligned}
    \bm{a}_1  &=   \frac{ \bm{e}^{\circ}_{IP_i} \bm{e}_{IP_i}^\top \bm{e}_{IG} }{d_{IP_i}} - \frac{\alpha_{ij}  \bm{e}^{\circ}_{IE_j} \bm{e}_{IE_j}^\top \bm{e}_{IG} }{d_{IE_j}}  - \frac{ d_f a_2 \bm{e}_{IG}^{\circ} }{1 + d_{IG} a_2} \\
    A & =  \frac{\alpha_{ij} d_f \bm{e}_{IG}^\circ  \bm{e}_{IG}^\top \bm{e}^{\circ}_{IE_j} \bm{e}_{IE_j}^\top}{d_{IE_j}}, \ a_2 =  \frac{\bm{e}_{IG}^{\circ \top} H(\bm{x}_G) \bm{e}_{IG}^{\circ}}{\| g_{\bm{y}}(\bm{x}_G) \|_2}.
\end{aligned}
\end{equation}

Proof. \rm
{\revise We prove the theorem by first proving that $\mathcal{X}^{\textup{ps}}_{ij}$ in \eqref{eq:position-initial-ERP} is a set of ERP winning states, where the key is to show that \eqref{eq:position-initial-ERP} is closed under the strategy \eqref{eq:sink-strategy-1v1}. Additionally, we spend much space to obtain the parameter condition \eqref{eq:parameters-1v1} to ensure that the strategy \eqref{eq:sink-strategy-1v1} is feasible, i.e., $|u_{P_i}| \leq 1$. Finally, we prove that the states meeting \eqref{eq:configuration-1v1} can be generated by the ERP winning states in $\mathcal{X}^{\textup{ps}}_{ij}$, via the steer-to-ERP approach in \thomref{thom:nece-suff-no-closing-nv1}, provided that \eqref{eq:parameters-1v1} holds.}

For simplicity, the subscripts $i$ and $j$ will be omitted in the proof. We first prove that $\mathcal{X}^{\textup{ps}}$ in \eqref{eq:position-initial-ERP} is a set of ERP winning states. 

We first prove that $\theta_P \equiv \sigma(X)$ under the strategy \eqref{eq:sink-strategy-1v1} if it holds initially. By the definition of $\sigma$ {\revise above \eqref{eq:convex-pbm-AC-PEF}} and the dynamics \eqref{eq:pursuer_car}, this equivalently implies that
\begin{equation}\label{eq:intercept-angle-1v1}
    \dot{\bm{x}}_P \equiv v_P \bm{e}_{IP},
\end{equation}
{\revise i.e., $P$'s heading is always pointing at the point $\bm{x}_I$.} Thus, we only need to verify \eqref{eq:intercept-angle-1v1} under the strategy \eqref{eq:sink-strategy-1v1} given it holds initially. Taking the time derivative for \eqref{eq:intercept-angle-1v1}, we have
\begin{equation}
\begin{aligned}\label{eq:stay-on-sink-proof-1v1}
    & \dod{\dot{\bm{x}}_P}{t} - v_P \dod{\bm{e}_{IP}}{t} 
    \\
    & = - \dot{\bm{x}}_{P}^\circ \dot{\theta}_P - v_P \frac{\dot{\bm{x}}_I - \dot{\bm{x}}_P - \bm{e}_{IP} \bm{e}_{IP}^\top (\dot{\bm{x}}_I - \dot{\bm{x}}_P) }{ d_{IP}}\\
     & = v_P \bm{e}_{IP}^\circ \frac{v_P}{\kappa} \frac{\kappa}{v_P} \frac{\bm{e}_{IP}^{\circ \top} \dot{\bm{x}}_I}{d_{IP}} - v_P \frac{\dot{\bm{x}}_I - \bm{e}_{IP} \bm{e}_{IP}^\top \dot{\bm{x}}_I}{ d_{IP}} = 0,
\end{aligned}
\end{equation}
where {\revise the first equality follows from the definitions of $\dot{\bm{x}}_P$ and $\bm{e}_{IP}$}, and the second equality follows from \eqref{eq:intercept-angle-1v1}, the dynamics \eqref{eq:pursuer_car} and the strategy \eqref{eq:sink-strategy-1v1} for $\theta_P = \sigma(X)$. Next, we show that if $X \in \mathcal{X}^{\textup{ps}}$, then $\varrho(X^t; f^{\textup{ps}}) > 0$ for all $t \ge 0$. It suffices to prove that the speed of $\mathbb{E}(X; f^{\textup{ps}})$ moving away from $\goal$ is non-negative, i.e., $\od{}{t}\varrho(X; f^{\textup{ps}}) \ge 0$, for all $X \in \mathcal{X}^{\textup{ps}}$. According to \eqref{eq:varrho-to-f-nv1} and \eqref{eq:f-to-control-nv1}, we have
\begin{equation*}
\begin{aligned}
    &\dod{}{t}\varrho(X; f^{\textup{ps}}) =  \dod{}{t}d  ( \bm{x}_I,\bm{x}_G) \\
    & = \lambda ( f^{\textup{ps}\top}_{P} (\bm{x}_I, X) \dot{\bm{x}}_P +  v_E f_{E}^{\textup{ps}\top} (\bm{x}_I, X) \bm{u}_E ) \\
    & = \lambda ( - v_P + v_E \alpha \bm{e}_{IE}^\top \bm{u}_E) \ge \lambda ( - v_P + v_E \alpha ) = 0,
\end{aligned}
\end{equation*}
where $\lambda \leq 0$ is the Lagrange multiplier,  $f^{\textup{ps}}_{P} = \partial f^{\textup{ps}} / \partial \bm{x}_P$ and $f^{\textup{ps}}_{E} = \partial f^{\textup{ps}} / \partial \bm{x}_E$, {\revise and the third equality follows from \eqref{eq:positional-PEF} and \eqref{eq:intercept-angle-1v1}}.
From the above, $X^t \in \mathcal{X}^{\textup{ps}}$ for all $t \ge 0$ and all $X \in \mathcal{X}^{\textup{ps}}$, under the strategy \eqref{eq:sink-strategy-1v1}.

In order to implement the strategy \eqref{eq:sink-strategy-1v1} for $\theta_P = \sigma(X)$, we need to ensure $|u_P| \leq 1$. To that end, we present the computation of $\dot{\bm{x}}_I$. Note that $(\bm{x}_I, \bm{x}_G)$ satisfies the KKT conditions \eqref{eq:kkt-condition-nv1} consistently. For \eqref{itm:winning-KKT-2}, we have
\begin{equation}\label{eq:parallel-gy-IG}
\begin{aligned}
    & g_{\bm{y}}(\bm{x}_G) \, || \, (\bm{x}_I - \bm{x}_G) \Rightarrow g^{\top}_{\bm{y}}(\bm{x}_G) (\bm{x}_I - \bm{x}_G)^{\circ} \equiv 0 \\
    & \Rightarrow \dod{}{t} \big( g^{\top}_{\bm{y}}(\bm{x}_G) (\bm{x}_I - \bm{x}_G)^{\circ}  \big ) = 0  \Rightarrow \\
    &g_{\bm{y}}^{\circ\top} (\bm{x}_G) \dot{\bm{x}}_I = ( H^\top(\bm{x}_G) (\bm{x}_I - \bm{x}_G)^{\circ}  + g^{\circ}_{\bm{y}} (\bm{x}_G))^\top \dot{\bm{x}}_G,
\end{aligned}    
\end{equation}
where $||$ denotes the parallel of two vectors. For \eqref{itm:winning-KKT-4}, we have
\begin{equation}\label{eq:stay-at-g}
\begin{aligned}
    & g(\bm{x}_G) \equiv 0 \Rightarrow \dod{}{t} g(\bm{x}_G) = 0 \Rightarrow g_{\bm{y}}^\top (\bm{x}_G) \dot{\bm{x}}_G = 0 .
\end{aligned}    
\end{equation}
By \eqref{eq:parallel-gy-IG} and \eqref{eq:stay-at-g}, $\dot{\bm{x}}_G$ is computed by
\begin{equation}\label{eq:dot-A-GI}
\begin{aligned}
     \dot{\bm{x}}_G & =  \frac{g^{\circ}_{\bm{y}}(\bm{x}_G) g_{\bm{y}}^{\circ \top} (\bm{x}_G) \dot{\bm{x}_I}}{\| g_{\bm{y}}(\bm{x}_G) \|_2^2 + (\bm{x}_I - \bm{x}_G)^{\circ \top} H(\bm{x}_G) g^{\circ}_{\bm{y}}(\bm{x}_G)} \\
     & =  \frac{\| g_{\bm{y}}(\bm{x}_G)\|_2^2 \bm{e}_{IG}^{\circ} \bm{e}^{\circ \top}_{IG} \dot{\bm{x}_I}}{\| g_{\bm{y}}(\bm{x}_G) \|_2^2 + d_{IG} \| g_{\bm{y}}(\bm{x}_G) \|_2 \bm{e}_{IG}^{\circ \top} H(\bm{x}_G) \bm{e}_{IG}^{\circ}} \\
    & = \frac{ \bm{e}_{IG}^{\circ} \bm{e}^{\circ \top}_{IG} \dot{\bm{x}_I}}{1 + d_{IG} a_2},
\end{aligned}
\end{equation}
where $a_2$ is defined in \eqref{eq:b1-B-1v1}, {\revise and the second equality is due to the fact that  $g^{\circ}_{\bm{y}}(\bm{x}_G) = \| g_{\bm{y}}(\bm{x}_G) \|_2 \bm{e}^{\circ}_{IG}$ using \eqref{eq:parallel-gy-IG} and \eqref{itm:winning-KKT-4}.} By combining  \eqref{itm:winning-KKT-1} and \eqref{itm:winning-KKT-2}, we can obtain that  $f^{\textup{ps}}_{ \bm{x}}(\bm{x}_I, X) \, || \, g_{\bm{y}}(\bm{x}_G)$, i.e., $f^{\textup{ps}\top }_{ \bm{x}}(\bm{x}_I, X) g_{\bm{y}}^{\circ}(\bm{x}_G) \equiv 0 $, where $f^{\textup{ps}}_{\bm{x}} = \partial f^{\textup{ps}} / \partial \bm{x}$. Thus $\od{}{t}  ( f^{\textup{ps}\top }_{ \bm{x}}(\bm{x}_I, X) g^{\circ}_{\bm{y}}(\bm{x}_G) ) = 0$ and thus
\begin{equation}\label{eq:parallel-fx-gy}
\begin{aligned}
    \bm{c}^\top_{1} \dot{\bm{x}}_I + \bm{c}^\top_2 \dot{\bm{x}}_P +  \bm{c}^\top_{3} \dot{\bm{x}}_E + \bm{c}^\top_4  \dot{\bm{x}}_{G}  = 0,
\end{aligned}    
\end{equation}
where
\begin{equation*}
\begin{aligned}
    \bm{c}_1 & =   \frac{ \bm{e}^{\circ}_{IP} \bm{e}_{IP}^\top \bm{e}_{IG} }{d_{IP}} - \frac{\alpha  \bm{e}^{\circ}_{IE} \bm{e}_{IE}^\top \bm{e}_{IG} }{d_{IE}}, \ \bm{c}_2 = -  \frac{ \bm{e}^{\circ}_{IP} \bm{e}_{IP}^\top \bm{e}_{IG} }{d_{IP}} \\
    \bm{c}_3 & =  \frac{\alpha  \bm{e}^{\circ}_{IE} \bm{e}_{IE}^\top \bm{e}_{IG} }{d_{IE}}, \  \bm{c}_4 = - \frac{d_f  H^\top (\bm{x}_G) \bm{e}_{IG}^{\circ}}{\| g_{\bm{y}}(\bm{x}_G)\|_2}.
\end{aligned}
\end{equation*}
{\revise In order to obtain \eqref{eq:parallel-fx-gy},  \eqref{eq:positional-PEF} and $g^{\circ}_{\bm{y}}(\bm{x}_G) = \| g_{\bm{y}}(\bm{x}_G) \|_2 \bm{e}^{\circ}_{IG}$ are both used.}
For \eqref{itm:winning-KKT-3}, we have $f^{\textup{ps}}(\bm{x}_I, X) \equiv 0$, that is, $\od{}{t} f^{\textup{ps}}(\bm{x}_I, X) = 0$, leading to
\begin{equation}\label{eq:stay-at-f}
\begin{aligned}
d_f \bm{e}^\top_{IG} \dot{\bm{x}}_I - \bm{e}_{IP}^\top \dot{\bm{x}}_P +  \alpha \bm{e}^\top_{IE} \dot{\bm{x}}_E = 0,
\end{aligned}    
\end{equation}
{\revise where \eqref{itm:winning-KKT-1} is used, i.e., $f^{\textup{ps}}_{ \bm{x}}(\bm{x}_I, X) = d_f \bm{e}_{IG}$, and $d_{f} = \| \bm{e}_{IP} - \alpha \bm{e}_{IE} \|_2$.} By substituting \eqref{eq:dot-A-GI} into \eqref{eq:parallel-fx-gy} and combining \eqref{eq:stay-at-f}, $\dot{\bm{x}}_I$ satisfies
\begin{equation}
\begin{bmatrix}
\bm{c}_1^\top + \bm{c}_5^\top \\
d_f \bm{e}^\top_{IG}
\end{bmatrix}
\dot{\bm{x}}_I + 
\begin{bmatrix}
\bm{c}^\top_2 \dot{\bm{x}}_P + \bm{c}_3^\top \dot{\bm{x}}_E \\
\alpha \bm{e}^\top_{IE} \dot{\bm{x}}_E - \bm{e}_{IP}^\top \dot{\bm{x}}_P
\end{bmatrix}
 = \bm{0},
\end{equation}
from which
\begin{equation}\label{eq:dot-xI}
\dot{\bm{x}}_I = \frac{ k_1 (\bm{c}_1  +  \bm{c}_5)^{\circ} - k_2
d_f \bm{e}_{IG}^{\circ}
}{
k_3
},
\end{equation}
where 
\begin{equation*}
\begin{aligned}
    k_1 & = \alpha \bm{e}^\top_{IE} \dot{\bm{x}}_E -\bm{e}_{IP}^\top \dot{\bm{x}}_P, &  k_2 & = \bm{c}^\top_2 \dot{\bm{x}}_P + \bm{c}_3^{\top} \dot{\bm{x}}_E \\
    k_3 & = d_f (\bm{c}_1 + \bm{c}_5)^\top  \bm{e}_{IG}^{\circ},  & \bm{c}_5 & = - \frac{ d_f a_2 \bm{e}_{IG}^{\circ} }{1 + d_{IG} a_2},
\end{aligned}
\end{equation*}
{\revise where the definition of $a_2$ in \eqref{eq:b1-B-1v1} is used.}
Note that
\begin{equation}\label{eq:k8-k9-bound}
   |k_1| \leq 2 v_P, \quad | k_2 | \leq v_P / d_{IP} + v_P / d_{IE}.
\end{equation}
Let $\gamma_{PG} = \bm{e}_{IP}^\top \bm{e}_{IG}$ and $\gamma_{EG} = \bm{e}_{IE}^\top \bm{e}_{IG}$, then $k_3$ can be rewritten as follows
\begin{equation}\label{eq:k3}
\begin{aligned}
    k_3 = \frac{d_f}{ d_{IP}} \big (\gamma_{PG}^2 - \frac{\alpha d_{IP}  \gamma_{EG}^2}{d_{IE}} \big ) - d_f \| \bm{c}_5 \|_2.
\end{aligned}
\end{equation}
{\revise Since $f^{\textup{ps}}_{ \bm{x}}(\bm{x}_I, X) = d_f \bm{e}_{IG}$,} this implies that
\begin{equation}\label{eq:fx-gy-same-direction}
\begin{aligned}
   d_f = (\bm{e}_{IP} - \alpha \bm{e}_{IE})^\top \bm{e}_{IG} = \gamma_{PG} - \alpha \gamma_{EG}.
\end{aligned}
\end{equation}
Since $d_f \in [\alpha - 1, \alpha + 1]$ by definition and $f^{\textup{ps}}(\bm{x}_I, X) = 0$ leads to $d_{IP} = \alpha d_{IE} + r$, by combining them with \eqref{eq:k3} and \eqref{eq:fx-gy-same-direction} we have the inequality
\begin{equation}\label{eq:k3-strictly-negative}
\begin{aligned}
    & \gamma_{PG}^2 - \frac{\alpha d_{IP}  \gamma_{EG}^2}{d_{IE}} = \gamma_{PG}^2 - \frac{(\alpha d_{IE} + r)(d_f - \gamma_{PG})^2}{\alpha d_{IE}} \\
    & = \frac{- r \gamma_{PG}^2 + 2 d_f (\alpha d_{IE} + r )\gamma_{PG}  - (\alpha d_{IE} + r ) d_f^2 }{\alpha  d_{IE}} \\
    & \leq - \frac{r (d_f - 1)^2 + \alpha d_f d_{IE}(d_f - 2)}{\alpha  d_{IE}} \coloneqq - (r k_4 / d_{IE} + k_5),
\end{aligned}
\end{equation}
where $\alpha > 3$ by assumption in \eqref{eq:parameters-1v1} is used in the last  inequality which follows when $\gamma_{PG} = 1$, and
\begin{equation}
    k_4 = (d_f - 1)^2 / \alpha, \quad k_5 = d_f (d_f - 2).
\end{equation}
Since $k_4, k_5 > 0$, then \eqref{eq:k3-strictly-negative} implies that $|k_3|$ {\revise in \eqref{eq:k3}} has a positive lower bound
\begin{equation}\label{eq:k3-lower-bound}
    |k_3| \ge  (r k_4 / d_{IE} + k_5) d_f / d_{IP} + d_f \| \bm{c}_5 \|_2.
\end{equation}
Using \eqref{eq:dot-xI} and the bounds for $k_1$, $k_2$ and $k_3$ in \eqref{eq:k8-k9-bound} and \eqref{eq:k3-lower-bound}, we can derive an upper bound for $\| \dot{\bm{x}}_I \|_2$ as follows
\begin{equation}\label{eq:up-xI-first}
\begin{aligned}
  &\| \dot{\bm{x}}_I \|_2 \leq \frac{2 v_P (\| \bm{c}_1\|_2 + \| \bm{c}_5 \|_2 ) + d_f (v_P / d_{IP} + v_P / d_{IE})}{ (r k_4 / d_{IE} + k_5) d_f / d_{IP} + d_f \| \bm{c}_5 \|_2}  \\
  & = \frac{
  2 v_Pd_{IP} (\| \bm{c}_1\|_2 + \| \bm{c}_5 \|_2 ) + d_f v_P + d_f d_{IP} v_P / d_{IE}
  }{ (r k_4 / d_{IE} + k_5) d_f + d_{IP}  d_f \| \bm{c}_5 \|_2 } \\
  & \leq \frac{2 v_P}{d_f} + \frac{ r k_6 / (\alpha d_{IE}) + k_7 }{ (r k_4 / d_{IE} + k_5) d_f + d_{IP}  d_f \| \bm{c}_5 \|_2} v_P, 
\end{aligned}
\end{equation}
where $k_6$ and $k_7$ are respectively given by
\begin{equation*}
\begin{aligned}
k_6 & = -2 d_f^2 + (\alpha + 4) d_f + 2 \alpha^2 - 2 \ge \alpha (\alpha + 1) > 0 \\
k_7 & = -2 d_f^2 + (\alpha + 5) d_f + 2 \alpha^2 + 2 \ge k_6 > 0,
\end{aligned}
\end{equation*}
{\revise In \eqref{eq:up-xI-first}, $\|\bm{c}_1\|_2 \leq 1/ d_{IP} + \alpha / d_{IE}$ and $d_{IP} = \alpha d_{IE} + r$ are used, and $k_6 \ge \alpha (\alpha + 1)$ is due to $d_f \in [\alpha - 1, \alpha + 1]$.} Then, the upper bound is further derived as follows
\begin{equation}\label{eq:dot-xi-1}
\begin{aligned}
     \| \dot{\bm{x}}_I \|_2 & \leq \frac{2 v_P}{d_f} + \frac{ r k_6 / (\alpha d_{IE}) + k_7 }{ (r k_4 / d_{IE} + k_5) d_f } v_P \\
     & \leq \frac{2 v_P}{\alpha - 1} +  v_P \max \{ k_6 / (\alpha d_f k_4), k_7 / ( d_fk_5 ) \},
\end{aligned}
\end{equation}
where 
\begin{equation*}
\begin{aligned}
    \frac{k_6}{ \alpha d_f k_4} & = \frac{2 \alpha^2 }{ d_f (d_f - 1)^2} + \frac{\alpha}{ (d_f - 1)^2} - \frac{2}{ d_f } \\
    & \leq \frac{2 \alpha^2 }{(\alpha - 1) (\alpha - 2)^2} + \frac{ \alpha }{(\alpha - 2)^2} - \frac{2}{ \alpha + 1 } \\
    & = \frac{ \alpha^3 + 12 \alpha^2 - 17 \alpha + 8}{(\alpha^2 - 1) (\alpha - 2 )^2}
\end{aligned}
\end{equation*}
\begin{equation*}
\begin{aligned}
    \frac{k_7}{ d_f k_5 } & = \frac{2 \alpha^2 + 2}{d_f^2 (d_f -2 )} + \frac{\alpha + 5}{ d_f (d_f - 2)} - \frac{2}{ d_f - 2} \\
    & \leq \frac{2 \alpha^2 + 2}{(\alpha - 1)^2 (\alpha - 3)} + \frac{\alpha + 5}{ (\alpha - 1) (\alpha - 3)} - \frac{2}{\alpha - 1} \\
    & = \frac{ \alpha^2 + 12 \alpha - 9}{(\alpha - 1)^2 (\alpha -3)}.
\end{aligned}
\end{equation*}
{\revise where $d_f \in [\alpha - 1, \alpha + 1]$ and the definitions of $k_4, k_5, k_6$ and $k_7$ are used.} Note that the upper bound of $k_7 / ( d_f k_5)$ is larger than the upper bound of $k_6 / (\alpha d_f k_4)$. Thus, the upper bound \eqref{eq:dot-xi-1} becomes
\begin{equation}\label{eq:dot-xi-bound-1v1}
\begin{aligned}
    \| \dot{\bm{x}}_I \|_2 \leq \frac{ 3 \alpha^2 + 4 \alpha - 3  }{(\alpha - 1)^2 (\alpha -3)} v_P.
\end{aligned}
\end{equation}
By combining \eqref{eq:dot-xi-bound-1v1} and \eqref{eq:sink-strategy-1v1}, the control $u_P$ for $\theta_P = \sigma(X)$ has the following bound
\begin{equation*}
  |u_P| = \big | - \frac{\kappa}{v_P} \frac{\bm{e}_{IP}^{\circ \top} \dot{\bm{x}}_I}{ d_{IP} } \big | \leq \frac{\kappa \| \dot{\bm{x}}_I \|_2}{v_P r} \leq \frac{ \kappa (3 \alpha^2 + 4 \alpha - 3) }{r (\alpha - 1)^2 (\alpha -3)},
\end{equation*}
{\revise where $d_{IP} \ge r$ is used}, which implies that $|u_P| \leq 1$ holds under the parameter condition \eqref{eq:parameters-1v1}. Thus, $\mathcal{X}^{\textup{ps}}$ is a set of ERP winning states.

By \thomref{thom:nece-suff-no-closing-nv1}, we next prove that the states meeting~\eqref{eq:configuration-1v1} can be generated by the ERP winning states in $\mathcal{X}^{\textup{ps}}$, via the steer-to-ERP approach. To that end, we compute such a time horizon $T$ in \thomref{thom:nece-suff-no-closing-nv1} via a lower bound of the angle chasing speed $|\dot{\theta}_P| - |\dot{\sigma}|$. Since the speed of the heading function $\sigma$ is bounded by
\begin{equation}
\begin{aligned}\label{eq:sigma-dot-bound-1v1}
    | \dot{\sigma} | & \leq \frac{\| \dot{\bm{x}}_I \|_2 + \| \dot{\bm{x}}_P \|_2 }{d_{IP}} \leq \frac{\| \dot{\bm{x}}_I \|_2 + v_P }{ r },
\end{aligned} 
\end{equation}
then under the strategy \eqref{eq:sink-strategy-1v1} for $\theta_P \neq \sigma(X)$ we have 
\begin{equation}\label{eq:angle-speed-diff-1v1}
\begin{aligned}
    | \dot{\theta}_P| - | \dot{\sigma} | & \ge v_P / \kappa - (\| \dot{\bm{x}}_I \|_2 + v_P ) / r \\
    & \ge \frac{v_P}{ r } \big ( \frac{r}{ \kappa } - 1 - \frac{ 3 \alpha^2 + 4 \alpha - 3}{(\alpha - 1)^2 (\alpha -3)} \big ) > 0,
\end{aligned}
\end{equation}
where \eqref{eq:dot-xi-bound-1v1} and \eqref{eq:parameters-1v1} are used. Since the maximum value of $|\theta_P - \sigma|$ to reduce under \eqref{eq:sink-strategy-1v1} for $\theta_P \neq \sigma(X)$ is $\pi $, any initial state will meet the condition $\theta_P = \sigma(X)$ within at most time $T \coloneqq \pi / (| \dot{\theta}_P| - | \dot{\sigma} |)$. {\revise In order to ensure the positive safe distance before $\theta_P = \sigma(X)$, we need to compute the speed $\od{}{t}{\varrho} (X; f^{\textup{ps}})$ of $\mathbb{E}(X; f^{\textup{ps}})$ moving away from $\goal$.} The KKT condition \eqref{itm:winning-KKT-1} implies that $|\lambda| = 1 / d_f$. {\revise By \thomref{thom:nece-suff-no-closing-nv1} and \rekref{rek:steer-to-ERP-speed}, $\od{}{t}{\varrho} (X; f^{\textup{ps}})$ is the optimal value of \eqref{eq:optimization-pbm-set-nv1} and is bounded by
\begin{equation}
\begin{aligned}\label{eq:rho-dist-1v1-final}
    \dod{}{t}{\varrho} (X; f^{\textup{ps}}) & \ge - |\lambda | (v_P |f_P^{\textup{ps} \top} \bm{e}_{\theta_P} |  + v_E \| f_E^{\textup{ps}} \|_2 ) \\
    & \ge - (v_P + v_E \alpha ) / d_f  \ge 2 v_P / (1 - \alpha).
\end{aligned}
\end{equation}
}
Therefore, using \eqref{eq:angle-speed-diff-1v1} and \eqref{eq:rho-dist-1v1-final},  if a state $X \in \mathcal{S}$ satisfies
\begin{equation}
    \varrho(X; f^{\textup{ps}}) >  \frac{ 2\pi r }{ (\alpha - 1) (r / \kappa - \textup{CM}_1(\alpha))},
\end{equation}
then {\revise $\varrho(X; f^{\textup{ps}}) + T \min_{t' \in [0, T] }\od{}{t}{\varrho} (X^{t'}; f^{\textup{ps}}) > 0$}, implying that $\varrho(X^{t'}; f^{\textup{ps}}) > 0$ for all $t'  \in [0, t^{\star}]$ and $X^{t^{\star}}$ is an ERP winning state in $\mathcal{X}^{\textup{ps}}$ for some $t^{\star} \in [0, T]$.
\qed
\end{thom} 

{\revise
\begin{rek}
    By \eqref{eq:sink-strategy-1v1}, \eqref{eq:intercept-angle-1v1} and \eqref{eq:stay-on-sink-proof-1v1}, in order to ensure that $\theta_{P_i} = \sigma_i(X_{ij})$ holds once reached, the current control input of $E_j$ is required to compute $\dot{\bm{x}}_I$ as it involves the term $\dot{\bm{x}}_{E_j}$. In practice, if the methods in Section~\ref{subsec:information-structure} are used to estimate or measure $\dot{\bm{x}}_{E_j}$ and the errors are not negligible, then we need to relax $\theta_{P_i} = \sigma_i(X_{ij})$ based on the error bounds to generalise $\mathcal{X}_{ij}^{\textup{ps}}$ in \eqref{eq:position-initial-ERP}, for which a possibly looser winning condition and a robust strategy might be obtained. Since the related robustness analysis is not straightforward, we leave it for future work.
\end{rek}
}

The proof of \thomref{thom:sink-winning-AC-1v1} shows that $\mathcal{X}_{ij}^{\textup{ps}}$ is a set of ERP winning states under the parameters \eqref{eq:parameters-1v1} which can be relaxed further as below.

\begin{lema}{\rm (Relaxed ERP winning parameters).}\label{lema:relaxed-ERP-win-1v1}
If the state satisfies $X_{ij} \in \mathcal{X}_{ij}^{\textup{ps}}$ and the parameters satisfy
\begin{equation}\label{eq:on-sink-paras-1v1}
    \alpha_{ij} > 3, \ r_i / \kappa_i \ge \frac{ (5 \alpha_{ij}^2 - 9 \alpha_{ij} + 8) }{ (\alpha_{ij} - 1) (\alpha_{ij} -2 )^2}
\end{equation}
then $X_{ij}$ is an ERP winning state and the strategy \eqref{eq:sink-strategy-1v1} is an ERP winning strategy.

\rm Proof. For simplicity, the subscripts $i$ and $j$ will be omitted in the proof. Since $X \in \mathcal{X}^{\textup{ps}}$, then \eqref{eq:intercept-angle-1v1} holds and thus the control $u_P$ for $\theta_P = \sigma(X)$ in \eqref{eq:sink-strategy-1v1} satisfies
\begin{equation}
\begin{aligned}
  & |u_P| = \big | \frac{\kappa}{v_P} \frac{\bm{e}_{IP}^{\circ \top} \dot{\bm{x}}_I}{ d_{IP} } \big |  = \frac{
  \kappa | k_1 \bm{e}_{IP}^\top ( \bm{c}_5 - \bm{c}_3 ) - k_2 d_f \bm{e}_{IP}^\top \bm{e}_{IG}| 
  }{v_P d_{IP} |k_3| } \\
  & \leq \frac{
  \kappa (2v_P (\| \bm{c}_5 \|_2 + \alpha / d_{IE}) + v_P d_f / d_{IE} )
  }{v_P ((r k_4 / d_{IE} + k_5) d_f+ d_f d_{IP}  \| \bm{c}_5 \|_2) }\\
  & = \frac{2 \kappa}{d_f d_{IP}} + \frac{\kappa ( (2\alpha +  d_f ) / d_{IE} - 2 (r k_4 / d_{IE} + k_5) / d_{IP} )
  }{(r k_4 / d_{IE} + k_5) d_f+ d_f d_{IP}  \| \bm{c}_5 \|_2 } \\
  & \leq \frac{2 \kappa}{d_f d_{IP}} + \frac{\kappa  (2\alpha +  d_f ) / d_{IE} 
  }{(r k_4 / d_{IE} + k_5) d_f } \leq  \frac{2 \kappa}{d_f r} + \frac{\kappa (2 \alpha + d_f)}{ r k_4 d_f } \\
  & = \frac{\kappa}{ r } \big ( \frac{2}{ d_f} + \frac{\alpha (2 \alpha + d_f)}{ d_f (d_f - 1)^2} \big) \leq \frac{ \kappa (5 \alpha^2 - 9 \alpha + 8) }{ r (\alpha - 1) (\alpha -2 )^2} \leq 1,
\end{aligned}
\end{equation}
where the second equality is due to \eqref{eq:dot-xI}, 
the first inequality follows from \eqref{eq:k8-k9-bound}, \eqref{eq:k3-lower-bound} and $\bm{e}_{IP}^{\circ\top} \dot{\bm{x}}_P = 0$ by \eqref{eq:intercept-angle-1v1}, the third inequality is based on the fact that $d_{IP} \ge r$ and $d_{IE} > 0$, the fourth inequality is because the formulation is monotonically decreasing in $d_f$ which has the range $[\alpha - 1, \alpha  + 1]$, and the last inequality holds using the condition \eqref{eq:on-sink-paras-1v1}. Therefore, the conclusion follows by the argument in the proof of \thomref{thom:sink-winning-AC-1v1}.
\qed
\end{lema}

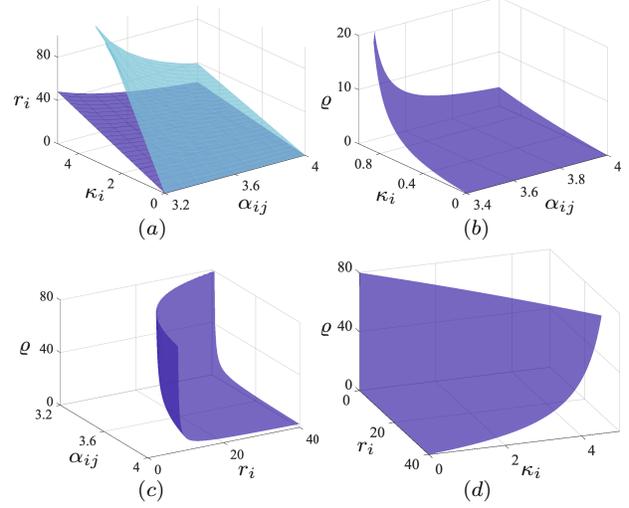
\begin{figure}
    \centering
    \includegraphics[width=0.45\hsize]{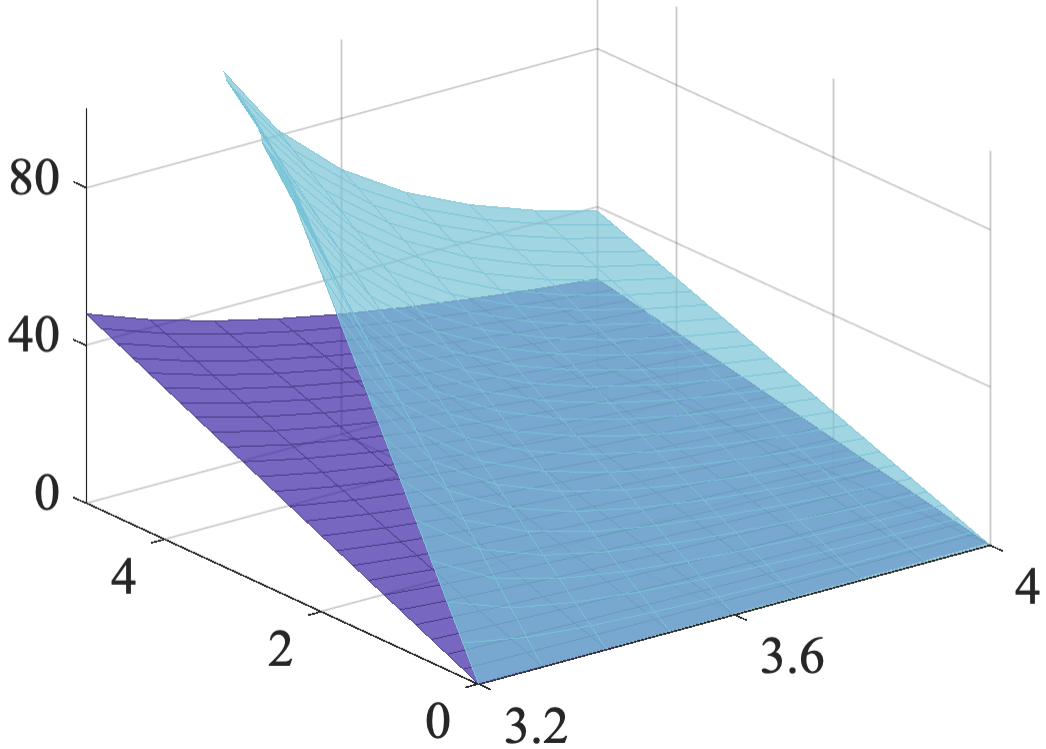}  \ \includegraphics[width=0.45\hsize]{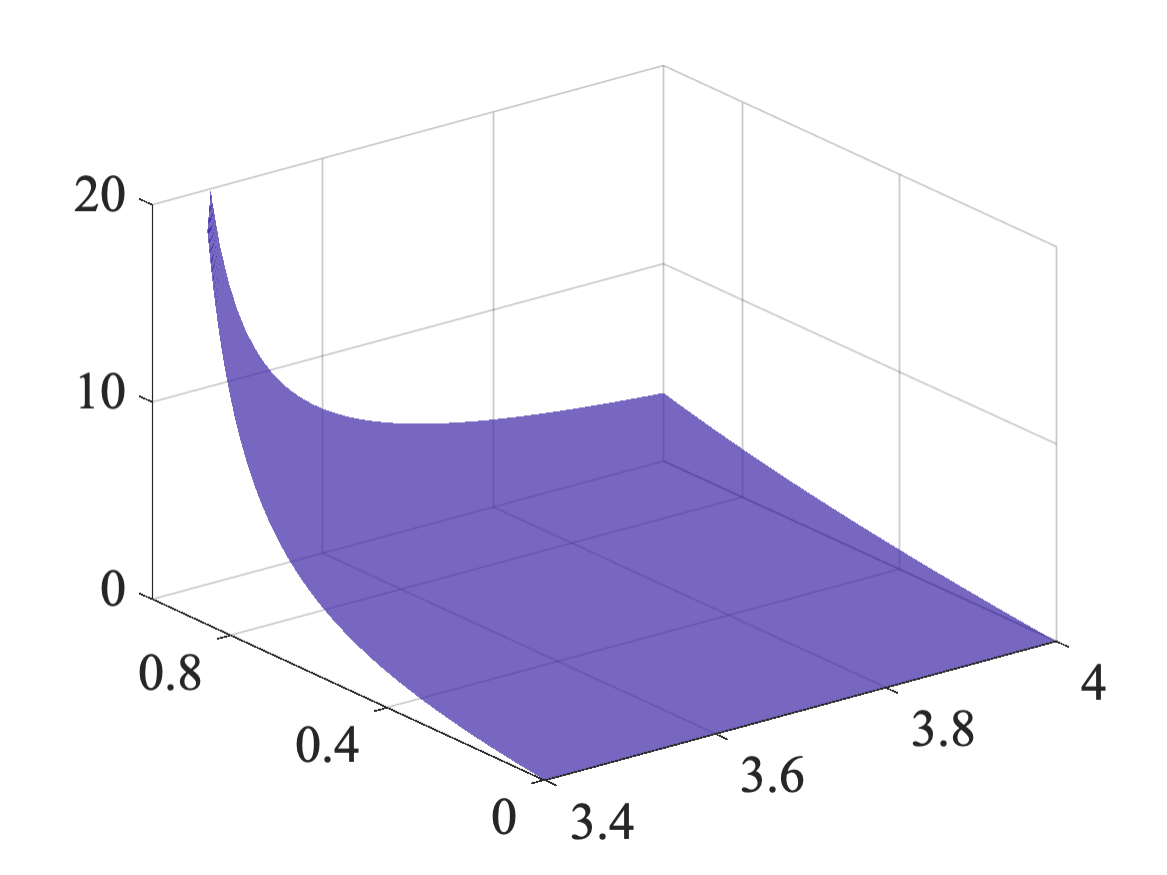}
    \put(-230,40){\scriptsize$r_i$}
    \put(-114,40){\scriptsize$\varrho$}
    \put(-203,5){\scriptsize$\kappa_i$}
    \put(-145,1){\scriptsize$\alpha_{ij}$}
    \put(-93,4){\scriptsize$\kappa_i$}
    \put(-29,1){\scriptsize$\alpha_{ij}$}
    \put(-60,-9){\scriptsize$(b)$}
    \put(-183,-9){\scriptsize$(a)$}
    \vspace{2pt}
    
    \includegraphics[width=0.45\hsize]{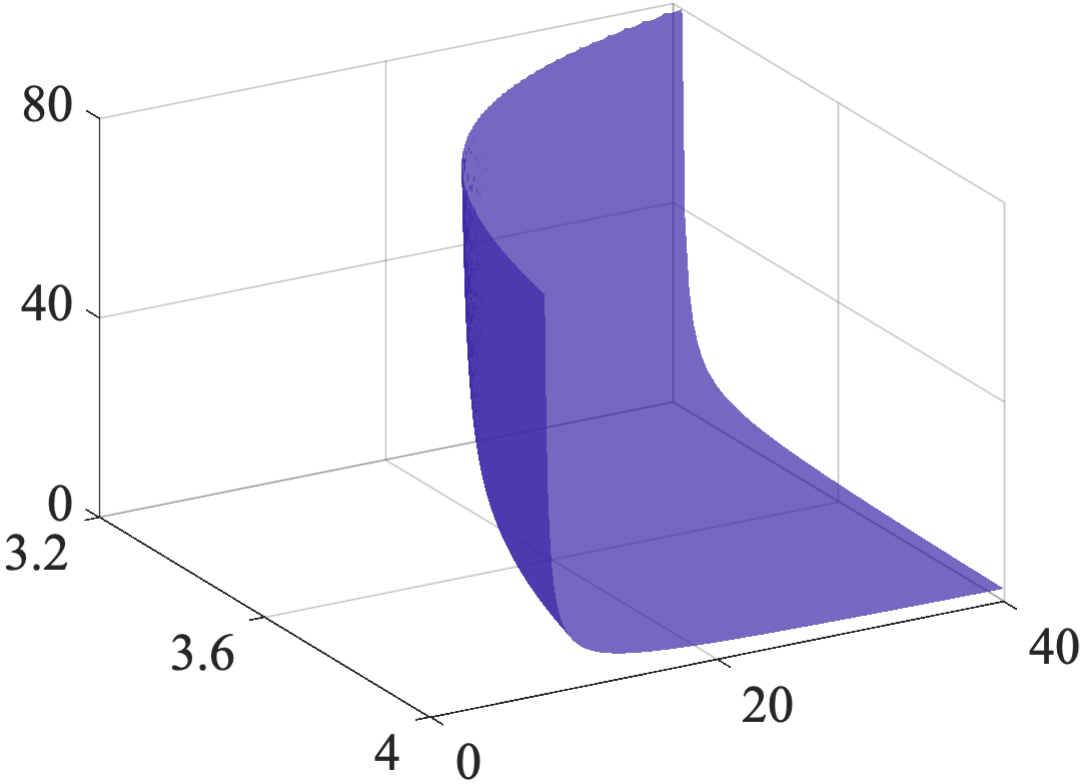} \
    \includegraphics[width=0.45\hsize]{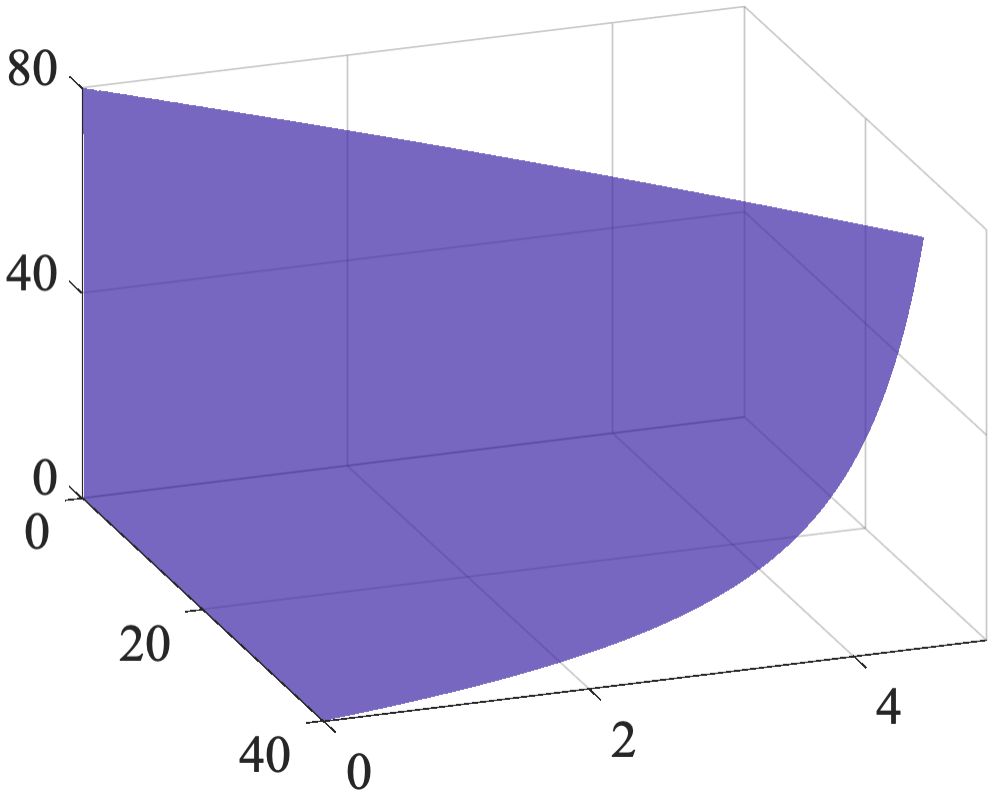}
    \put(-60,-8){\scriptsize$(d)$}
    \put(-183,-8){\scriptsize$(c)$}
    \put(-228,47){\scriptsize$\varrho$}
    \put(-209,5){\scriptsize$\alpha_{ij}$}
    \put(-145,0){\scriptsize$r_i$}
    \put(-114,54){\scriptsize$\varrho$}
    \put(-100,9){\scriptsize$r_i$}
    \put(-38,0){\scriptsize$\kappa_i$}
    \caption{The ERP winning conditions for one pursuer and one evader with the PEF \eqref{eq:positional-PEF}. $(a)$ parameters in \eqref{eq:parameters-1v1} (purple) and \eqref{eq:on-sink-paras-1v1} (blue); safe distance in \eqref{eq:configuration-1v1}: $(b)$  $r_i = 20$, $(c)$ $\kappa_i = 1$, $(d)$ $\alpha_{ij} = 4$.}
    \label{fig:winning-conditions-1v1}
\end{figure}

{\revise The ERP winning conditions in \thomref{thom:sink-winning-AC-1v1} and \lemaref{lema:relaxed-ERP-win-1v1}  for one pursuer and one evader with the PEF \eqref{eq:positional-PEF} are shown in Fig.~\ref{fig:winning-conditions-1v1}. In Fig.~\ref{fig:winning-conditions-1v1}$(a)$, the boundaries of winning parameters (i.e., $\alpha_{ij}, r_i$ and $ \kappa_i$) given by \eqref{eq:parameters-1v1} and \eqref{eq:on-sink-paras-1v1} are depicted in purple and blue, respectively. The boundary of the winning condition \eqref{eq:configuration-1v1} on the safe distance and parameters is depicted in Fig.~\ref{fig:winning-conditions-1v1}$(b)$-$(d)$ when the value of one of the parameters is fixed.}

\subsection{ERP winning conditions for two pursuers}

We next consider two pursuers and one evader. Based on $\mathcal{X}_{cj}^{\textup{ps}}$ and results of the one-pursuer case, we present the conditions on parameters and states that can ensure the ERP winning via the steer-to-ERP approach, and give the corresponding ERP winning strategies.

\begin{thom}\label{thom:sink-winning-AC-2v1}{\rm (ERP winning parameters, state and strategy).}
Let $c = \{ 1, 2 \}$. Consider a two-pursuer pursuit coalition $P_c = \{ P_1, P_2 \} $ against an evader $E_j$. 
There are two scenarios for $X_{cj} \in \mathcal{S}_{cj}$: 
\begin{enumerate}
    \item if $\mathcal{P}^{\textup{ps}}(X_{cj})$ has the unique support constraint (say $P_1$), then it goes to one-pursuer case in \thomref{thom:sink-winning-AC-1v1}. Verify $P_1$ against $E_j$ first, and if it fails, then verify $P_2$ against $E_j$;
    
    \item if $\mathcal{P}^{\textup{ps}}(X_{cj})$ has two support constraints, then: if the parameters \eqref{eq:parameters-1v1} hold for $P_1$ and $P_2$ against $E_j$ separately and the safe distance of $X_{cj}$ satisfies
    \begin{equation}\label{eq:erp-win-safe-distance}
    \varrho(X_{cj}; f_c^{\textup{ps}} ) > \min_{i' \in c} \max_{i \in c} \frac{2 \pi_i r_i v_{P_{i'}} / (\alpha_{i'}v_{P_i}  - v_{P_i} )}{r_i / \kappa_i - \textup{CM}_2(\alpha_1, \alpha_2)},
    \end{equation}
    where
    \begin{equation*}
    \begin{aligned}
        \textup{CM}_2 (\alpha_1, \alpha_2) = 1 + \min_{i''\in c}\frac{ 3 \alpha_{i''}^2 + 4 \alpha_{i''} - 3}{(\alpha_{i''} - 1)^2 (\alpha_{i''} -3)},
    \end{aligned}
    \end{equation*}
    then $X_{cj}$ is an ERP winning state and the cooperative feedback pursuit strategy
    \begin{equation}\label{eq:sink-strategy-2v1}
    u_{P_i} = 
    \begin{cases}
    - \frac{\kappa_i}{v_{P_i}} \frac{\bm{e}_{IP_i}^{\circ \top} \dot{\bm{x}}_I}{d_{IP_i} }, \ \textup{if } \theta_{P_i} = \sigma_i (X_{cj})\\
    \textup{sgn} (\sin (\sigma_i (X_{cj}) - \theta_{P_i})), \ \textup{otherwise } 
    \end{cases}
    \end{equation}
    for all $i \in c$ and all $X_{cj} \in \mathcal{S}_{cj}$, is an ERP winning strategy with $(\bm{x}_I, \bm{x}_G)$ computed by the convex problem \eqref{eq:convex-pbm-AC-PEF} and $\dot{\bm{x}}_I$ is given by
\begin{equation*}
    \frac{
    \bm{a}_{1}^\circ (\alpha_2 \bm{e}_{IE_j}^\top \dot{\bm{x}}_{E_j} -v_{P_2}) - \bm{a}_{2}^\circ (\alpha_1 \bm{e}_{IE_j}^\top \dot{\bm{x}}_{E_j} - v_{P_1})
    }{\bm{a}_{1}^\top \bm{a}_{2}^\circ }
\end{equation*}
where $\bm{a}_{i} = \bm{e}_{IP_i} - \alpha_{ij} \bm{e}_{IE_j}$ for $i \in c$.
\end{enumerate}

Proof. \rm
{\revise For the second scenario, similar to the proof of \thomref{thom:sink-winning-AC-1v1}, we first prove that $\mathcal{X}^{\textup{ps}}_{cj}$ in \eqref{eq:position-initial-ERP} is a set of ERP winning states by showing that \eqref{eq:position-initial-ERP} is closed under the strategy \eqref{eq:sink-strategy-2v1}. Then we also show that the strategy \eqref{eq:sink-strategy-2v1} is feasible, i.e., $|u_{P_i}| \leq 1$ for all $i \in c$ under the parameter conditions \eqref{eq:parameters-1v1}. Finally, we prove that the states meeting \eqref{eq:erp-win-safe-distance} can be generated by the ERP winning states in $\mathcal{X}^{\textup{ps}}_{cj}$, via the steer-to-ERP approach in \thomref{thom:nece-suff-no-closing-nv1}.}

Unless for clarity, the subscript $j$ for $E_j$, $\bm{x}_{E_j}, \bm{u}_{E_j}, v_{E_j}$, $\alpha_{ij}$ and $\bm{e}_{IE_j}$ is omitted in the proof. For the first scenario, since $\mathcal{P}^{\textup{ps}}(X_{cj})$ has the unique support constraint (say $P_1$), checking the case of $P_1$ against $E$ is sufficient. However, $P_2$ may be able to win against $E$ even if $P_1$ fails. Regarding the second scenario, we first prove that $\mathcal{X}^{\textup{ps}}_{cj}$ is a set of ERP winning states.

We first prove that $\theta_{P_i} \equiv \sigma_i(X_{cj})$ under the strategy~\eqref{eq:sink-strategy-2v1} if it holds initially. By the definition of $\sigma_i$ {\revise above \eqref{eq:convex-pbm-AC-PEF}} and the dynamics \eqref{eq:pursuer_car}, this equivalently implies that
\begin{equation}\label{eq:intercept-angle-2v1}
    \dot{\bm{x}}_{P_1} \equiv v_{P_1} \bm{e}_{IP_1}, \quad \dot{\bm{x}}_{P_2} \equiv v_{P_2} \bm{e}_{IP_2}.
\end{equation}
It can be proved by following the same argument in \eqref{eq:stay-on-sink-proof-1v1}. Next, we show that if $X_{cj} \in \mathcal{X}^{\textup{ps}}_{cj}$, then $\varrho(X_{cj}^t; f_c^{\textup{ps}}) > 0$ for all $t \ge 0$ under the strategy \eqref{eq:sink-strategy-2v1}. It suffices to prove that the speed of $\mathbb{E}(X_{cj}; f_c^{\textup{ps}})$ moving away from $\goal$ is non-negative, i.e., $\od{}{t}\varrho(X_{cj}; f_c^{\textup{ps}}) \ge 0$, for all $X_{cj} \in \mathcal{X}^{\textup{ps}}_{cj}$. According to \eqref{eq:varrho-to-f-nv1} and \eqref{eq:f-to-control-nv1}, we have
\begin{equation*}
\begin{aligned}
    &\dod{}{t}\varrho(X_{cj}; f_c^{\textup{ps}}) =  \dod{}{t} d ( \bm{x}_I,\bm{x}_G) \\
    & = \sum_{i \in c} \lambda_i ( f^{\textup{ps}\top}_{i, P} (\bm{x}_I, X_{ij}) \dot{\bm{x}}_{P_i} +  v_E f_{i,E}^{\textup{ps}\top} (\bm{x}_I, X_{ij}) \bm{u}_E ) \\
    & = \sum_{i \in c} \lambda_i (v_E \alpha_i \bm{e}_{IE}^\top \bm{u}_E - v_{P_i}) \ge \sum_{i \in c} \lambda_i (v_E \alpha_i - v_{P_i}) = 0.
\end{aligned}
\end{equation*}From the above, $X^t_{cj} \in \mathcal{X}^{\textup{ps}}_{cj}$ for all $t \ge 0$ and all $X_{cj} \in \mathcal{X}^{\textup{ps}}_{cj}$, under the strategy \eqref{eq:sink-strategy-2v1}.

Similar to the one-pursuer case, in order to implement the strategy \eqref{eq:sink-strategy-2v1} for $\theta_{P_i} = \sigma_i(X_{cj})$, we also need to ensure $|u_{P_i}| \leq 1$ for all $i \in c$. To that end, we present the computation of $\dot{\bm{x}}_I$ for two pursuers against one evader. Note that $(\bm{x}_I, \bm{x}_G)$ satisfies the KKT conditions \eqref{eq:kkt-condition-nv1} consistently. Since $\mathcal{P}^{\textup{ps}}(X_{cj})$ has two support constraints, then $\lambda_1 < 0$ and $\lambda_2 < 0$ ($c = \{ 1, 2\}$). Then, for \eqref{itm:winning-KKT-3}, we have
\begin{equation*}
    f_i^{\textup{ps}}(\bm{x}_I, X_{ij}) \equiv 0 \Rightarrow \dod{}{t} f_i^{\textup{ps}}(\bm{x}_I, X_{ij}) = 0 \textup{ for } i \in c,
\end{equation*}
that is,
\begin{equation*}
\begin{aligned}
\begin{bmatrix}
\bm{a}_{1}^\top \\
\bm{a}_{2}^\top
\end{bmatrix}
\dot{\bm{x}}_I + 
\begin{bmatrix}
\alpha_1 \bm{e}_{IE}^\top \dot{\bm{x}}_E - \bm{e}_{IP_1}^\top \dot{\bm{x}}_{P_1} \\
\alpha_2 \bm{e}_{IE}^\top \dot{\bm{x}}_E - \bm{e}_{IP_2}^\top \dot{\bm{x}}_{P_2}
\end{bmatrix}
 = \bm{0},
\end{aligned}
\end{equation*}
where $\bm{a}_{i} = \bm{e}_{IP_i} - \alpha_i \bm{e}_{IE}$ for $i \in c$. By solving the above equations, $\dot{\bm{x}}_I$ is given by
\begin{equation}\label{eq:dot-xI-2v1-ini}
    \frac{
    \bm{a}_{1}^\circ (\alpha_2 \bm{e}_{IE}^\top \dot{\bm{x}}_E - \bm{e}_{IP_2}^\top \dot{\bm{x}}_{P_2}) - \bm{a}_{2}^\circ (\alpha_1 \bm{e}_{IE}^\top \dot{\bm{x}}_E - \bm{e}_{IP_1}^\top \dot{\bm{x}}_{P_1} )
    }{\bm{a}_{1}^\top \bm{a}_{2}^\circ }.
\end{equation}
Since we only need to ensure $|u_{P_i}| \leq 1$ for $\theta_{P_i} = \sigma_i (X_{cj})$ and for all $i \in c$, i.e., \eqref{eq:intercept-angle-2v1} holds, then \eqref{eq:dot-xI-2v1-ini} is simplified as
\begin{equation}\label{eq:dot-xI-2v1}
    \dot{\bm{x}}_I = \frac{
    \bm{a}_{1}^\circ (\alpha_2 \bm{e}_{IE}^\top \dot{\bm{x}}_E -v_{P_2}) - \bm{a}_{2}^\circ (\alpha_1 \bm{e}_{IE}^\top \dot{\bm{x}}_E - v_{P_1})
    }{\bm{a}_{1}^\top \bm{a}_{2}^\circ }.
\end{equation}
Take $k_1 = \alpha_1 \bm{e}_{IE}^\top \dot{\bm{x}}_E - v_{P_1}$, $k_2 = \alpha_2 \bm{e}_{IE}^\top \dot{\bm{x}}_E -v_{P_2}$, $d_{1} = \| \bm{a}_{1} \|_2$, $d_{2} = \| \bm{a}_{2} \|_2$ and $\bm{a}_{1}^\top \bm{a}_{2} = d_{1} d_{2} \cos \gamma$, where $\gamma \in (0, \pi )$. Then we have
\begin{equation}\label{eq:dot-xi-bound-ini-2v1}
\begin{aligned}
    & \| \dot{\bm{x}}_I \|_2^2 = \frac{
    k_2^2 d_1^2 + k_1^2 d_2^2 - 2 k_1 k_2 d_1 d_2 \cos \gamma 
    }{ d_1^2 d_2^2 \sin^2 \gamma } \\
    & \Rightarrow \dod{}{\gamma } \| \dot{\bm{x}}_I \|_2^2 = \frac{2 (k_1 d_2 \cos \gamma - k_2 d_1) (k_2 d_1 \cos \gamma - k_1 d_2)}{ d_1^2 d_2^2 \sin^3 \gamma} \\
    & \Rightarrow \| \dot{\bm{x}}_I \|_2 \leq \max \big \{ \lim_{\gamma \to 0} \| \dot{\bm{x}}_I \|_2, \lim_{\gamma \to \pi} \| \dot{\bm{x}}_I \|_2 \big \}.  
\end{aligned}
\end{equation}
If $\gamma \to 0$, i.e., $\bm{a}_{1}$ is parallel to $\bm{a}_{2}$ with the same direction, then {\revise by \eqref{itm:winning-KKT-1}}, one enclosure region is an inscribed region of the other and thus it is degenerated into the one pursuer case. Then by \eqref{eq:dot-xi-bound-1v1}, we have
\begin{equation*}
    \lim_{\gamma \to 0} \| \dot{\bm{x}}_I \|_2 \leq \min_{i \in c} \frac{ 3 \alpha^2_i + 4 \alpha_i - 3  }{(\alpha_i - 1)^2 (\alpha_i -3)} v_{P_i}.
\end{equation*}
If $\gamma \to \pi$, i.e., $\bm{a}_{1}$ is parallel to $\bm{a}_{2}$ with the opposite direction, then two enclosure regions are externally tangent. This implies that $\bm{x}_I \to \bm{x}_E$, and thus we have
\begin{equation*}
    \lim_{\gamma \to \pi} \| \dot{\bm{x}}_I \|_2 = \| \dot{\bm{x}}_E \|_2 = v_E.
\end{equation*}
Therefore, $\| \dot{\bm{x}}_I \|_2 $ is bounded by
\begin{equation}\label{eq:xI-dot-2v1-final}
    \| \dot{\bm{x}}_I \|_2 \leq  \min_{i \in c} \frac{ 3 \alpha^2_i + 4 \alpha_i - 3  }{(\alpha_i - 1)^2 (\alpha_i -3)} v_{P_i}.
\end{equation}
By combining \eqref{eq:xI-dot-2v1-final} with \eqref{eq:sink-strategy-2v1}, the control $u_{P_i}$ for $\theta_{P_i} = \sigma_i (X_{cj})$ has the following bound
\begin{equation}\label{eq:u-p-upper-bound-2v1}
\begin{aligned}
  |u_{P_i}| & = \big | - \frac{\kappa_i}{v_{P_i}} \frac{\bm{e}_{IP_i}^{\circ \top} \dot{\bm{x}}_I}{ d_{IP_i} } \big | \leq  \frac{\kappa_i \| \dot{\bm{x}}_I \|_2 }{v_{P_i} r_i} \\
  & \leq \frac{\kappa_i }{v_{P_i} r_i}  \min_{i' \in c} \frac{ 3 \alpha^2_{i'} + 4 \alpha_{i'} - 3  }{(\alpha_{i'} - 1)^2 (\alpha_{i'} -3)} v_{P_{i'}} < 1,
\end{aligned}
\end{equation} 
where the parameter conditions \eqref{eq:parameters-1v1} are used. In conclusion, $\mathcal{X}^{\textup{ps}}_{cj}$ is a set of ERP winning states.

According to \thomref{thom:nece-suff-no-closing-nv1}, we next prove that the states meeting~\eqref{eq:erp-win-safe-distance} can be generated by the ERP winning states in $\mathcal{X}^{\textup{ps}}_{cj}$, via the steer-to-ERP approach. By following the same argument \eqref{eq:sigma-dot-bound-1v1} and \eqref{eq:angle-speed-diff-1v1}, the angle chasing speed has the positive lower bound
\begin{equation*}
\begin{aligned}
    |\dot{\theta}_{P_i}| - |\dot{\sigma}_i| {\ge} \frac{v_{P_i}}{ r_i } \big ( \frac{r_i}{ \kappa_i } - 1 - \min_{i'\in c}\frac{ v_{P_{i'}}(3 \alpha_{i'}^2 + 4 \alpha_{i'} - 3)}{v_{P_i}(\alpha_{i'} - 1)^2 (\alpha_{i'} -3)} \big ).
\end{aligned}
\end{equation*}
Thus, similar to the one-pursuer case, any initial state will meet the condition $\theta_{P_i} = \sigma_i(X_{cj})$ for all $i \in c$ within at most time $T \coloneqq \max_{i \in c} \pi / (| \dot{\theta}_{P_i}| - | \dot{\sigma}_i |)$. In order to ensure the positive safe distance before $\theta_{P_i} = \sigma_i(X_{cj})$ for all $i \in c$, we need to compute the speed $\od{}{t}{\varrho} (X_{cj}; f^{\textup{ps}}_c)$ of $\mathbb{E}(X_{cj}; f^{\textup{ps}}_c)$ moving away from $\goal$. By \thomref{thom:nece-suff-no-closing-nv1} and \rekref{rek:steer-to-ERP-speed}, $\od{}{t}{\varrho} (X_{cj}; f^{\textup{ps}}_c)$ is the optimal value of \eqref{eq:optimization-pbm-set-nv1} and is bounded by
\begin{equation}
\begin{aligned}
     & \dod{}{t}{\varrho} (X_{cj}; f^{\textup{ps}}_c)\\
    & \ge - \sum _{i\in c}|\lambda_i | (|f_{i,P}^\top \dot{\bm{x}}_{P_i} | + v_{P_i} |f_{i,\theta}|/ \kappa_i + v_E \| f_{i,E} \|_2 ) \\
    & \ge - (2 v_{P_1} |\lambda_1| + 2 v_{P_2} |\lambda_2|) = 2 v_{P_1} \lambda_1 + 2 v_{P_2} \lambda_2.
\end{aligned}
\end{equation}
Since $\lambda_1$ and $\lambda_2$ are subject to \eqref{itm:winning-KKT-1}, then we consider the optimization problem
\begin{equation}\label{eq:convex-pbm-vIG2}
\begin{aligned}
    & \underset{(\lambda_1,\lambda_2)\in\mathbb{R}^2} {\textup{minimize}}
	&& 2 v_{P_1} \lambda_1 + 2 v_{P_2} \lambda_2 \\
	&\textup{ subject to}&& \lambda_1 \bm{a}_{1} + \lambda_2 \bm{a}_{2} + \bm{e}_{IG} = \bm{0}.
\end{aligned}
\end{equation}
The KKT conditions to \eqref{eq:convex-pbm-vIG2} lead to
\begin{subequations}\label{eq:kkt-condition-vIG2}
\begin{align}
	& 2 v_{P_1} + \bm{z}^\top \bm{a}_{1} = 0, \quad  2 v_{P_2} + \bm{z}^\top \bm{a}_{2} = 0 \label{itm:winning-KKT-vIG2-1}\\
	& \lambda_1 \bm{a}_{1} + \lambda_2 \bm{a}_{2} + \bm{e}_{IG} = \bm{0}, \label{itm:winning-KKT-vIG2-2}
\end{align}
\end{subequations}
where $\bm{z} \in \mathbb{R}^2$ is the Lagrange multiplier, from which we have
\begin{equation}
\begin{aligned}
    2 v_{P_1} \lambda_1 + 2 v_{P_2} \lambda_2 & = - \lambda_1 \bm{z}^\top \bm{a}_{1} - \lambda_2 \bm{z}^\top \bm{a}_{2} \\
    & = \bm{z}^\top \bm{e}_{IG} \ge - \| \bm{z} \|_2,
\end{aligned}
\end{equation}
and
\begin{equation}\label{eq:z-expression-2v1}
    \bm{z} = (2v_{P_2} \bm{a}_{1}^\circ - 2v_{P_1} \bm{a}_{2}^\circ) / (\bm{a}_{1}^\top \bm{a}_{2}^\circ).
\end{equation}
Note that $\bm{z}$ has the similar expression as $\dot{\bm{x}}_I$ in \eqref{eq:dot-xI-2v1}. Then, following the argument \eqref{eq:dot-xi-bound-ini-2v1}, we obtain
\begin{equation}
    \| \bm{z} \|_2 \leq \max \big \{ \lim_{\gamma \to 0} \| \bm{z} \|_2, \lim_{\gamma \to \pi} \| \bm{z} \|_2 \big \}.
\end{equation}
If $\gamma \to 0$, it is degenerated into the one pursuer case, and if $\gamma \to \pi$, then $\bm{x}_I \to \bm{x}_E$. This observation implies that
\begin{equation}
    \od{}{t}{\varrho} (X_{cj}; f^{\textup{ps}}_c) \ge  \max_{i \in c} 2v_{P_i} / (1 - \alpha_i),
\end{equation}
where \eqref{eq:rho-dist-1v1-final} is used. Thus, if a state $X_{cj} \in \mathcal{S}_{cj}$ satisfies
\begin{equation*}
    \varrho(X_{cj}; f_c^{\textup{ps}}) > \min_{i' \in c} \max_{i \in c} \frac{2 \pi_i r_i v_{P_{i'}} / (\alpha_{i'}v_{P_i} - v_{P_i})}{ r_i / \kappa_i - \textup{CM}_2(\alpha_1, \alpha_2)},
\end{equation*}
then $\varrho(X_{cj}; f^{\textup{ps}}_c) + T \min_{t' \in [0, T] }\od{}{t}{\varrho} (X^{t'}_{cj}; f^{\textup{ps}}_c) > 0$, and thus $\varrho(X^{t'}_{cj}; f_c^{\textup{ps}}) > 0$ for $t'  \in [0, t^{\star}]$ and $X^{t^{\star}}_{cj}$ is an ERP winning state in $\mathcal{X}^{\textup{ps}}_{cj}$ for some $t^{\star} \in [0, T]$. \qed
\end{thom}

\begin{figure}
    \centering
    \includegraphics[width=0.45\hsize]{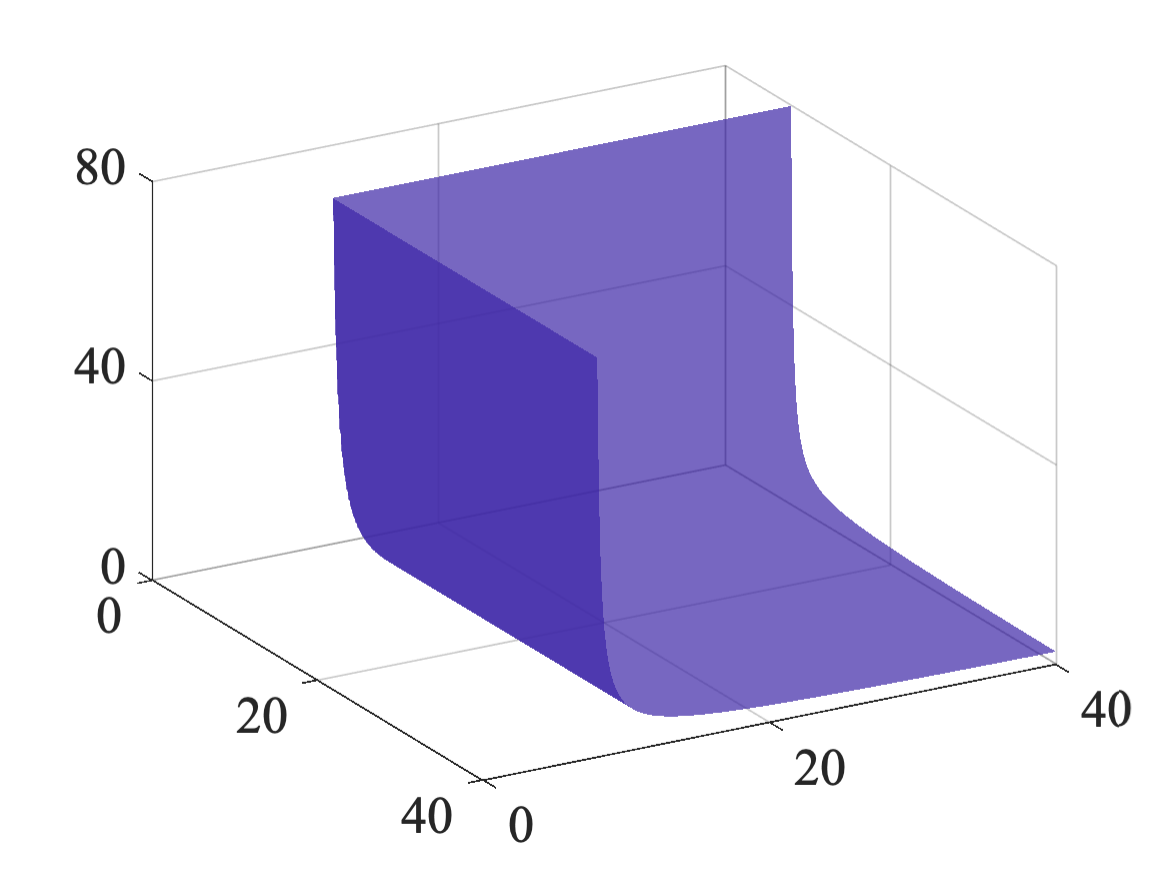} \
    \includegraphics[width=0.45\hsize]{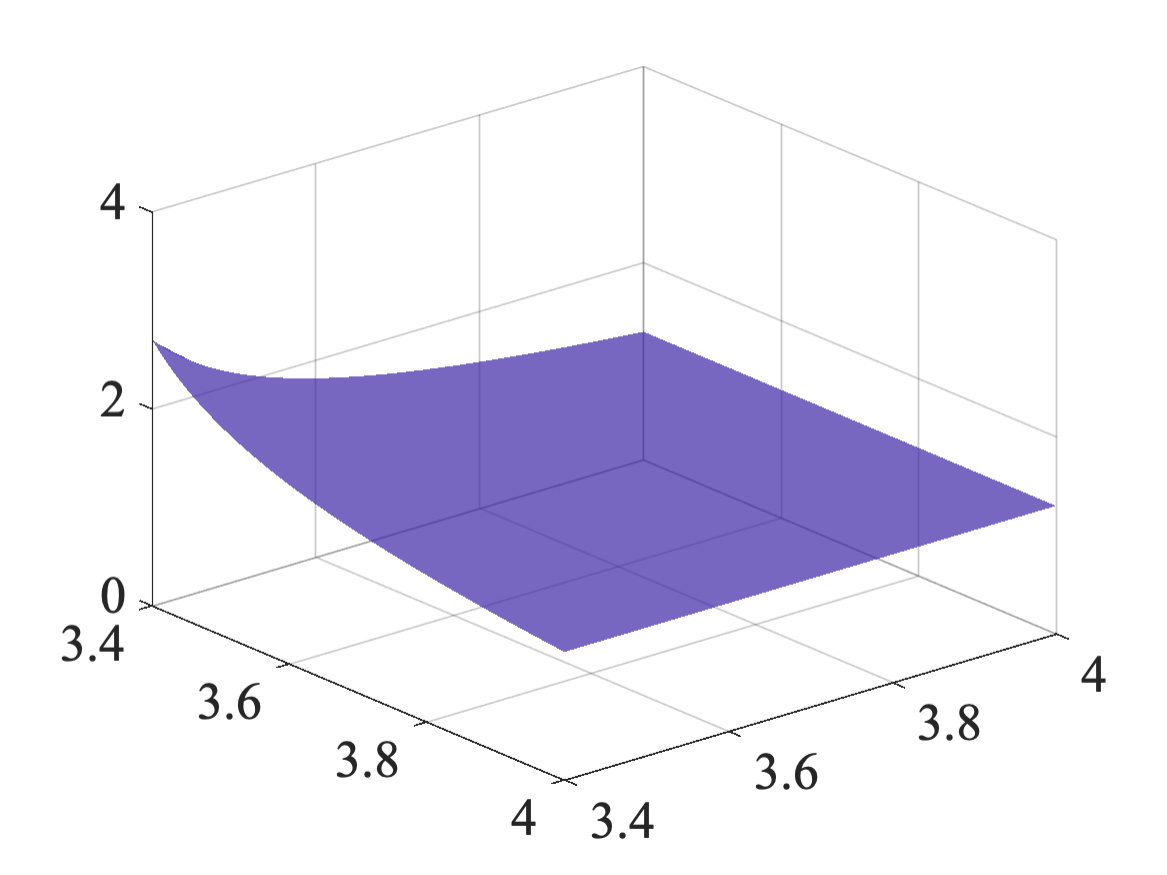}
    \put(-228,47){\scriptsize$\varrho$}
    \put(-209,5){\scriptsize$r_1$}
    \put(-145,0){\scriptsize$r_2$}
    \put(-60,-8){\scriptsize$(b)$}
    \put(-188,-8){\scriptsize$(a)$}
    \put(-110,44){\scriptsize$\varrho$}
    \put(-92,2){\scriptsize$\alpha_{1j}$}
    \put(-31,0){\scriptsize$\alpha_{2j}$}
    \caption{The ERP winning conditions for two pursuers and one evader with the PEF \eqref{eq:positional-PEF}. Safe distance in \eqref{eq:erp-win-safe-distance}: $(a)$ $\alpha_{1j} = \alpha_{2j} = 4$ and $\kappa_1 = \kappa_2 = 1$; $(b)$ $r_1 = r_2 = 20$ and $\kappa_1 = \kappa_2 = 0.5$.}
    \label{fig:winning-conditions-2v1}
\end{figure}
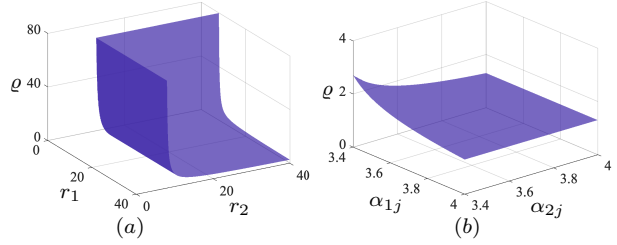

{\revise The ERP winning conditions in \thomref{thom:sink-winning-AC-2v1} for two pursuers and one evader with the PEF \eqref{eq:positional-PEF} are shown in Fig.~\ref{fig:winning-conditions-2v1}. Since the winning condition \eqref{eq:erp-win-safe-distance} on the safe distance and parameters involves seven variables,  we visualise the boundaries by fixing the values of speed ratios and minimum turning radii in Fig.~\ref{fig:winning-conditions-2v1}$(a)$, and capture radii and minimum turning radii in Fig.~\ref{fig:winning-conditions-2v1}$(b)$.}

\section{Pursuit Strategies based on Task Allocation}\label{sec:taks-allocation}

This section first considers the task allocation between pursuit coalitions and evaders by piecing together the outcomes of all subgames. Let $\mathcal{G} = (\mathcal{V}_P \cup \mathcal{V}_E, \mathcal{E})$ be an undirected bipartite graph consisting of two independent vertex sets $\mathcal{V}_P$ and $\mathcal{V}_E$, and a set of edges $\mathcal{E}$. In our problem, $\mathcal{V}_P$ is the set of all nonempty pursuit coalitions of size less than or equal to two, and $\mathcal{V}_E$ the set of evaders. The edge connecting vertex $P_c \in \mathcal{V}_P$ and vertex $E_j \in \mathcal{V}_E$ is denoted by $e_{cj}$, and $e_{cj} \in \mathcal{E}$ if and only if $P_c$ is able to defend against $E_j$ through the ERP winning. Let $\mathcal{C} = ( \mathcal{E}, \bar{\mathcal{E}})$ be a conflict graph, where each vertex in $\mathcal{C}$ corresponds uniquely to an edge in $\mathcal{G}$, and an edge $\bar{e} \in \bar{\mathcal{E}}$ if and only if two vertexes (two edges in $\mathcal{G}$) connecting by $\bar{e}$ involve at least one common pursuer.

We formulate the problem of maximizing the number of captured evaders in the ERP winning as a binary integer program (BIP):
\begin{equation}\label{eq:BIP}
\begin{aligned}
    & \underset{a_{cj}, a_{pq} \in \{ 0, 1\} }{\textup{maximize}}
	&& \sum\nolimits_{e_{cj} \in \mathcal{E} } a_{cj} + z(a_{cj})\\
	&\textup{ subject to}&& \sum\nolimits_{P_c \in \mathcal{V}_P } a_{cj} \leq 1, \quad \forall E_j \in \mathcal{V}_E \\
	& & & \sum\nolimits_{E_j \in \mathcal{V}_E} a_{cj} \leq 1, \quad \forall P_c \in \mathcal{V}_P \\
	& & & a_{cj} + a_{pq} \leq 1, \quad \forall (e_{cj}, e_{pq}) \in \bar{\mathcal{E}}
\end{aligned}
\end{equation}
where {\revise the subscripts $p$ and $q$ mean the pursuit coalition $P_p$ and evader $E_q$, respectively,} $a_{cj} = 1$ indicates the allocation of pursuit coalition $P_c$ to capture evader $E_j$, and $a_{cj} = 0$ means no assignment (similar for $a_{pq}$), and $z: \mathcal{E} \times \{0, 1\} \to \mathbb{R}$ evaluates the assignment. {\revise By the definition of $\mathcal{C}$, the last constraint in \eqref{eq:BIP} implies that for every edge in $\bar{\mathcal{E}}$, at most one associated assignment can be taken, which ensures that a pursuer does not get multiple assignments.} Let $L = 1+ \max_{P_c \in \mathcal{V}_P, E_j \in \mathcal{V}_E} \varrho(X_{cj}; f_c)$ be the maximum safe distance among players plus one.

\begin{thom}\label{thom:task-alloc}{\rm (Task allocation).}
For the BIP \eqref{eq:BIP},
\begin{enumerate}
    \item\label{itm:task-1} the complexity is NP-hard;
    
    \item\label{itm:task-2} if $z(a_{cj}) = 0$, the solution corresponds to the most captured evaders. 
    
    \item\label{itm:task-3} if $z(a_{cj}) = a_{cj} \varrho(X_{cj}; f_c) / (\min \{ N_p, N_e \}  L) $, the solution corresponds to the most captured evaders with the maximum sum of safe distances;
    
    \item\label{itm:task-4} if $z(a_{cj}) = - a_{cj} \varrho(X_{cj}; f_c) / (\min \{ N_p, N_e \}  L) $, the solution corresponds to the most captured evaders with the minimum sum of safe distances.
\end{enumerate}

Proof. \rm
Regarding (\ref{itm:task-1}), the identical argument to the proof of Theorem 4.1 in \cite{RY-XD-ZS-YZ-FB:19} proves that the well-known NP-complete \emph{3-dimensional matching problem} \citep{RMK:72} is polynomially reduced to special instances of the BIP \eqref{eq:BIP}. Regarding (\ref{itm:task-2}), it follows from the definition. Regarding (\ref{itm:task-3}), since $0 \leq \sum_{e_{cj} \in \mathcal{E}} z(a_{cj}) < \frac{\sum_{e_{cj} \in \mathcal{E}} a_{cj} L }{\min \{ N_p, N_e \}  L } \leq 1$, then $\sum_{e_{cj} \in \mathcal{E}} z(a_{cj})$ is strictly dominated by the value increment of matching one more evader. Thus, the conclusion follows directly, and (\ref{itm:task-4}) can be proved similarly.
\qed
\end{thom}

To solve \eqref{eq:BIP}, many solvers can be used (e.g., Gurobi, Matlab). If the number of players is large, the Sequential Matching Algorithm \citep{RY-XD-ZS-YZ-FB:19} is a $1/2$ approximation polynomial algorithm, and an exact algorithm if the solution does not contain pursuit coalitions with two pursuers. Motivated by \cite{RY-XD-ZS-YZ-FB:19,LA-JS-SG-JM:22}, we combine the ERP winning (motion planning) with the task allocation (task planning) in a receding-horizon manner and thus propose a multiplayer receding-horizon ERP strategy (\algoref{alg:multiplayer-re-ERP-stra}) that ensures a monotonically increasing number of guaranteed captured evaders.  Since at most two pursuers are needed for the ERP winning by the coalition reduction, the number of tasks to be considered is reduced from $(2^{N_p} - 1)N_e$ to $(N_p + 1) / N_p N_e$, {\revise  which however, as \thomref{thom:task-alloc} states, is still an NP-hard problem.} $\mathit{ERP\_Winning} (X_{cj}, f_c)$ determines whether $P_c$ guarantees an ERP winning against $E_j$ from $X_{cj}$. This can be verified using the proposed steer-to-ERP approach with the positional PEFs $\{ f_i^\textup{ps} \}_{P_i \in \pteam} $, i.e, check the winning parameters and safe distances in Theorems \ref{thom:sink-winning-AC-1v1} or \ref{thom:sink-winning-AC-2v1}.

\begin{algorithm}
\caption{Multiplayer ERP strategy }\label{alg:multiplayer-re-ERP-stra}
\textbf{Initialize:} $\{ \bm{x}_{P_i}, \theta_{P_i} \}_{P_i \in \pteam}$, $\{ \bm{x}_{E_j} \}_{E_j \in \eteam}$, PEFs $\{ f_i \}_{P_i \in \pteam}$

\begin{algorithmic}[1]
\State $\mathcal{V}_P \leftarrow \{ P_c \in 2^{\pteam } \mid 1\leq |P_c| \leq 2 \}$, $\mathcal{V}_E \leftarrow \eteam$
\Repeat
\For{$P_c \in \mathcal{V}_P$, $E_j \in \mathcal{V}_E$}
\State Add $e_{cj}$ to $\mathcal{E}$ if $\mathit{ERP\_Winning} (X_{cj}, f_c)$ is true
\EndFor

\State $G \leftarrow (\mathcal{V}_P \cup \mathcal{V}_E, \mathcal{E})$
\State $M \leftarrow $ solve the BIP \eqref{eq:BIP}
\State Adopt ERP winning strategy \eqref{eq:sink-strategy-1v1} or \eqref{eq:sink-strategy-2v1} for $P_c$ 

if $(P_c, E_j) \in M$ for some $E_j$
\State Adopt some (any) strategy for $E_j \in \mathcal{V}_E$ and 

unmatched $P_i \in \pteam$
\State Update $\bm{x}_{P_i}, \theta_{P_i}, \bm{x}_{E_j}$ with a time step $\Delta$
\State Remove captured or arriving evaders from $\mathcal{V}_E$
\Until{$\mathcal{V}_E = \emptyset $}
\end{algorithmic}
\end{algorithm}

\section{Simulations}\label{sec:simulation}

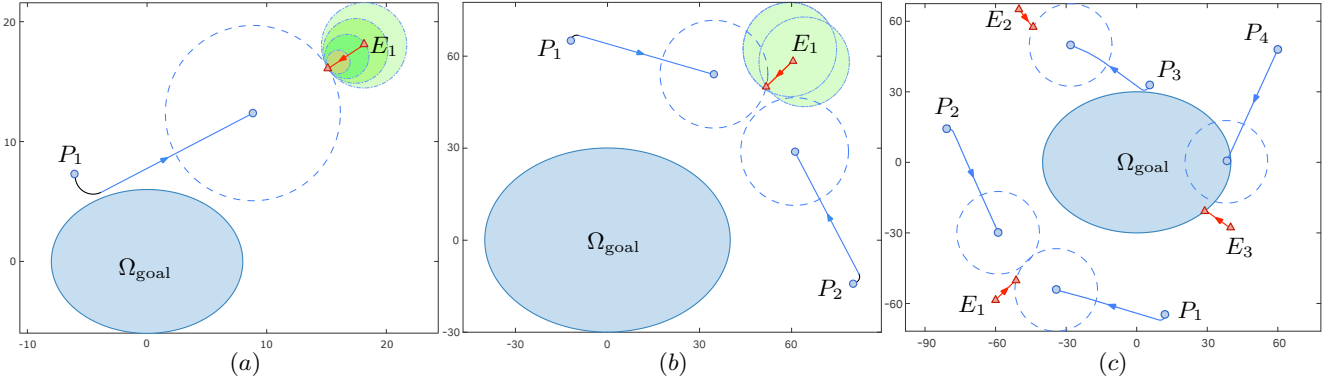
\begin{figure*}
    \centering
    \includegraphics[width=0.33\hsize]{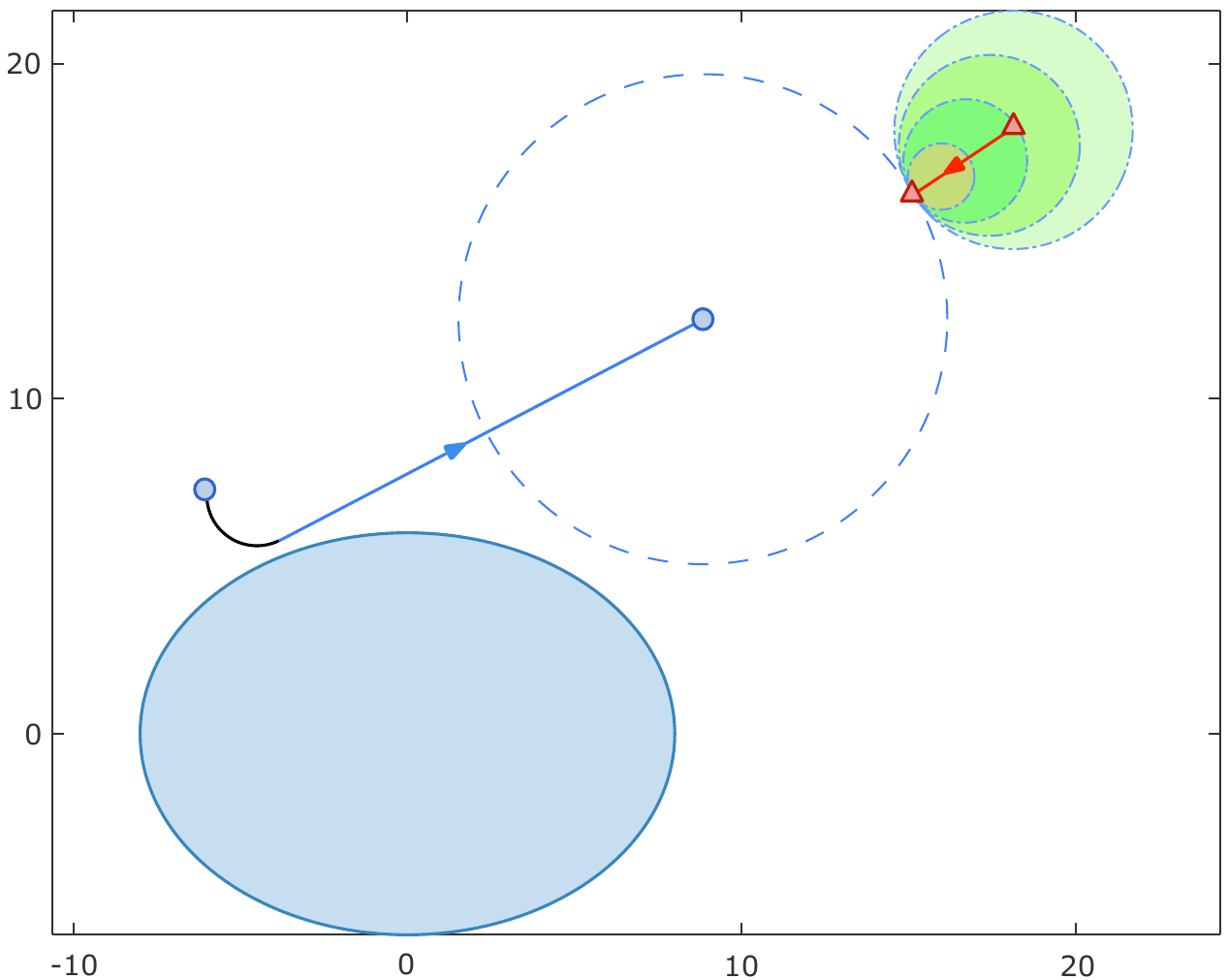}  
    \put(-122,28){\footnotesize$\goal$}
    \put(-145,71){\footnotesize$P_1$}
    \put(-27,112){\footnotesize$E_1$}
    \put(-80,-8){\footnotesize$(a)$}
    \includegraphics[width=0.332\hsize]{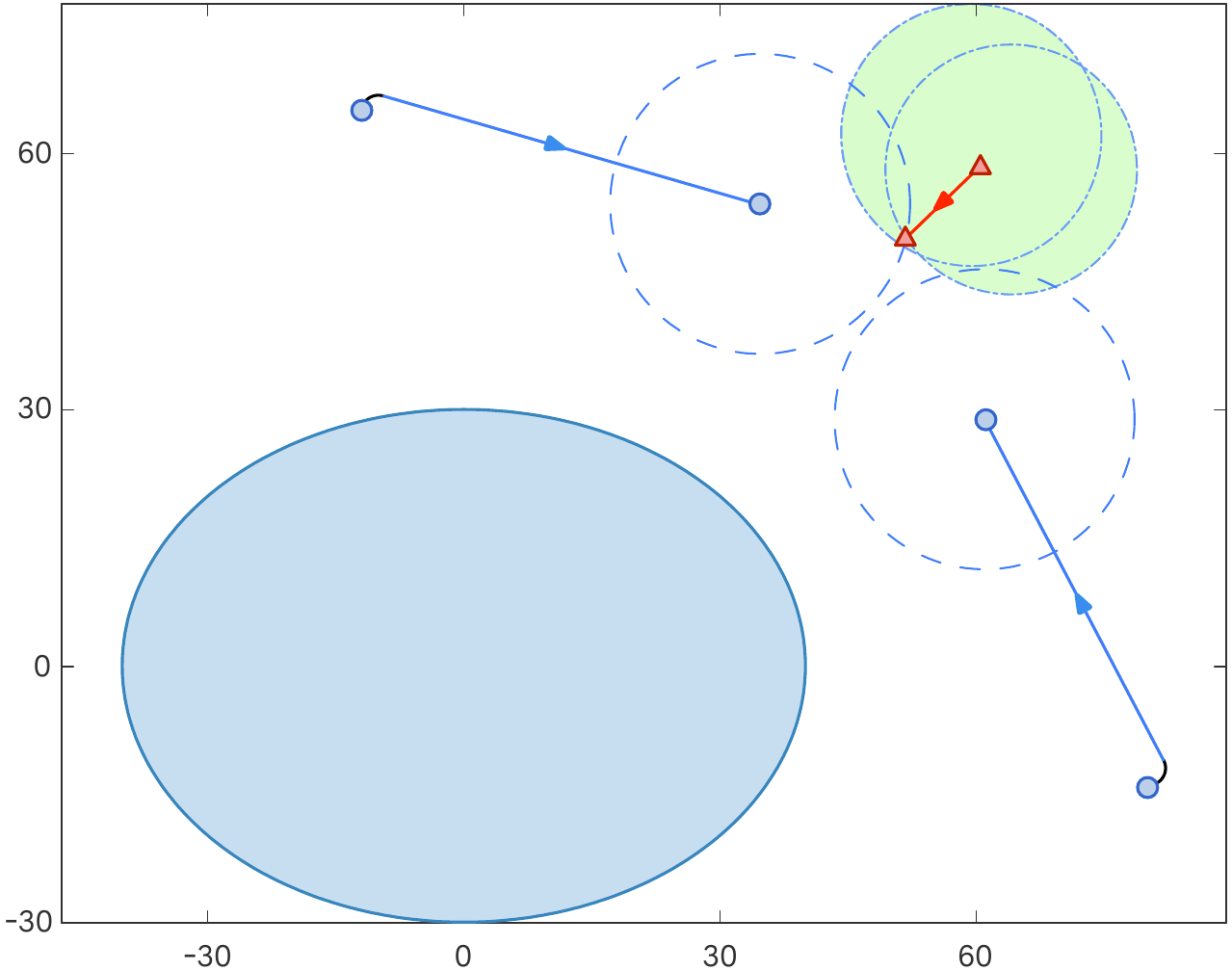} 
    \put(-112,38){\footnotesize$\goal$}
    \put(-131,111){\footnotesize$P_1$}
    \put(-25,20){\footnotesize$P_2$}
    \put(-35,113){\footnotesize$E_1$}
    \put(-85,-8){\footnotesize$(b)$}
    \includegraphics[width=0.33\hsize]{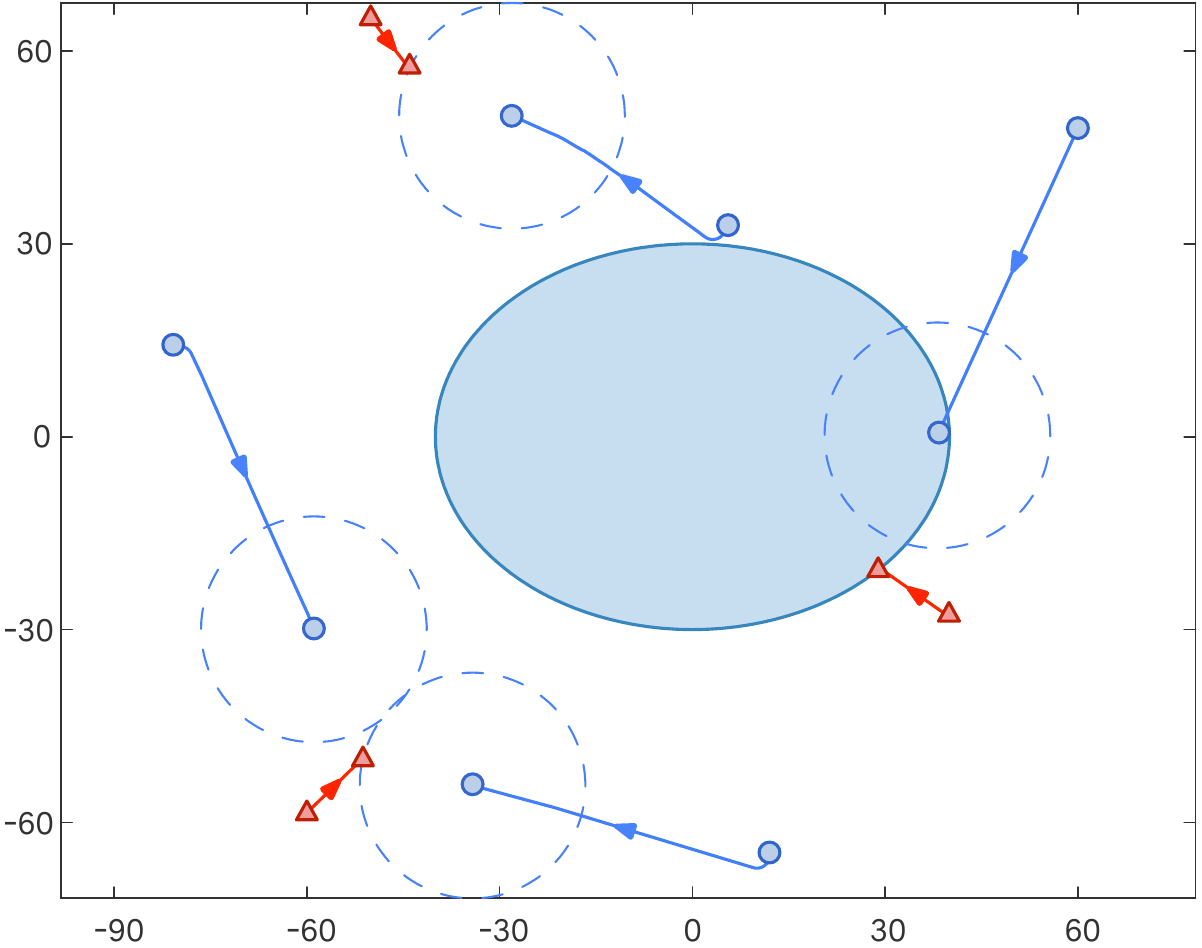} 
    \put(-78,68){\footnotesize$\goal$}
    \put(-55,12){\footnotesize$P_1$}
    \put(-147,89){\footnotesize$P_2$}
    \put(-63,103){\footnotesize$P_3$}
    \put(-30,117){\footnotesize$P_4$}
    \put(-138,14){\footnotesize$E_1$}
    \put(-37,36){\footnotesize$E_3$}
    \put(-128,122){\footnotesize$E_2$}
    \put(-85,-8){\footnotesize$(c)$}
    \caption{Three simulations. $(a)$ one pursuer and one evader; $(b)$ two pursuers and one evader; $(c)$ four pursuers and three evaders.}
    \label{fig:simulation}
\end{figure*}

We run the Homicidal Chauffeur reach-avoid differential games in various scenarios with different team sizes and initial configurations to illustrate the theoretical results. The positional PEF \eqref{eq:positional-PEF} is used for the ERP winning.

Case 1: one pursuer $P_1$ and one evader $E_1$. We consider the parameters {\revise $\alpha_{11} = 5, r_1 = 7.31$ and $\kappa_1 = 1.5$} which satisfy the condition \eqref{eq:parameters-1v1}, and consider the initial states $\bm{x}_{P_1} = [-6, 7], \theta_{P_1} = - 1.5$ and $\bm{x}_{E_1} = [18, 18]$ which satisfy the condition \eqref{eq:configuration-1v1}. By \thomref{thom:sink-winning-AC-1v1}, the initial state is an ERP winning state and thus under the ERP winning strategy \eqref{eq:sink-strategy-1v1}, $P_1$ is able to defend the goal region against $E_1$ which can take any strategy. The scenario is depicted in Fig. \ref{fig:simulation}$(a)$, where the blue dashed circle is the capture range. After the state enters $\mathcal{X}_{11}^{\textup{ps}}$ in \eqref{eq:position-initial-ERP}, the ERP winning strategy by $P_1$ ensures that the enclosure regions containing $E_1$ (in green at several time instants) never approach the goal region, and thus $E_1$ cannot reach the goal region before being captured.

Case 2: two pursuers $P_1, P_2$ and one evader $E_1$. We consider $\alpha_{11} = \alpha_{21} = 4$, $r_1 = r_2 =17.56$ and $\kappa_1 = \kappa_2 = 2$. The initial states are $\bm{x}_{P_1} = [-12, 65], \theta_{P_1} = 1.5$, $\bm{x}_{P_2} = [80, -14], \theta_{P_2} = -0.1$ and $\bm{x}_{E_1} = [60.5, 58.5]$. By \thomref{thom:sink-winning-AC-1v1}, $P_1$ and $P_2$ cannot ensure the ERP winning against $E_1$ individually due to the failure of the state condition \eqref{eq:configuration-1v1}. However, the initial states meet the condition  $(2)$ in \thomref{thom:sink-winning-AC-2v1}, and thus $P_1$ and $P_2$ can defend against $E_1$ by cooperation using the strategy \eqref{eq:sink-strategy-2v1}, as in Fig. \ref{fig:simulation}$(b)$, {\revise where the enclosure regions are depicted at the instant when the state enters $\mathcal{X}_{cj}^{\textup{ps}}$ in \eqref{eq:position-initial-ERP}.}

Case 3: four pursuers and three evaders. The multiplayer receding-horizon ERP strategy is used in this example. The task allocation shows that the pursuit team can ensure the  simultaneous ERP winning against at most two evaders, show in Fig. \ref{fig:simulation}$(c)$. More concretely, $P_1$ and $P_2$ cooperatively defend against $E_1$, and $P_3$ defends against $E_2$. The pursuer $P_4$ is tasked to pursue $E_3$, although it cannot ensure the ERP winning against $E_3$.

\section{Conclusion}\label{sec:conclusion}

We presented a cooperative pursuit strategy for multiplayer Homicidal Chauffeur reach-avoid differential games in which the pursuers protect a convex region against the evaders. For the subgames, the ERP winning provides a sufficient condition to guarantee the pursuit winning without directly working with the terminal conditions. In addition to avoiding the backward analysis, the ERP winning has simple cooperation among pursuers due to the pursuit coalition reduction. The steer-to-ERP approach shows that, if a set of ERP winning states are constructed, the new ERP winning states can be generated by solving an optimization problem. The parameters, states and strategies that ensure the ERP winning with the proposed positional PEFs are able to find a part of the pursuit winning conditions. The task allocation leads to an increasing number of guaranteed captured evaders. Future work will involve two-car dynamics and distributed games. {\revise Moreover, since the task allocation at each step involves solving a combinatorial problem, for future work, we will propose heuristic methods based on players' current states to prune the pursuit coalitions that are considered for possible tasks.}

 \bibliographystyle{model5-names}
\bibliography{reference}           

\begin{thebibliography}{49}
\expandafter\ifx\csname natexlab\endcsname\relax\def\natexlab#1{#1}\fi
\providecommand{\url}[1]{\texttt{#1}}
\providecommand{\href}[2]{#2}
\providecommand{\path}[1]{#1}
\providecommand{\DOIprefix}{doi:}
\providecommand{\ArXivprefix}{arXiv:}
\providecommand{\URLprefix}{URL: }
\providecommand{\Pubmedprefix}{pmid:}
\providecommand{\doi}[1]{\href{http://dx.doi.org/#1}{\path{#1}}}
\providecommand{\Pubmed}[1]{\href{pmid:#1}{\path{#1}}}
\providecommand{\bibinfo}[2]{#2}
\ifx\xfnm\relax \def\xfnm[#1]{\unskip,\space#1}\fi
\bibitem[{Antonyshyn et~al.(2023)Antonyshyn, Silveira, Givigi \&
  Marshall}]{LA-JS-SG-JM:22}
\bibinfo{author}{Antonyshyn, L.}, \bibinfo{author}{Silveira, J.},
  \bibinfo{author}{Givigi, S.}, \& \bibinfo{author}{Marshall, J.}
  (\bibinfo{year}{2023}).
\newblock \bibinfo{title}{Multiple mobile robot task and motion planning: A
  survey}.
\newblock {\it \bibinfo{journal}{ACM Computing Surveys}\/},  {\it
  \bibinfo{volume}{55}\/}, \bibinfo{pages}{1--35}.
\bibitem[{Battistini \& Shima(2014)}]{SB-TS:14}
\bibinfo{author}{Battistini, S.}, \& \bibinfo{author}{Shima, T.}
  (\bibinfo{year}{2014}).
\newblock \bibinfo{title}{Differential games missile guidance with
  bearings-only measurements}.
\newblock {\it \bibinfo{journal}{IEEE Transactions on Aerospace and Electronic
  Systems}\/},  {\it \bibinfo{volume}{50}\/}, \bibinfo{pages}{2906--2915}.
\bibitem[{Bopardikar et~al.(2009)Bopardikar, Bullo \& Hespanha}]{SDB-FB-JPH:09}
\bibinfo{author}{Bopardikar, S.~D.}, \bibinfo{author}{Bullo, F.}, \&
  \bibinfo{author}{Hespanha, J.~P.} (\bibinfo{year}{2009}).
\newblock \bibinfo{title}{A cooperative {H}omicidal {C}hauffeur game}.
\newblock {\it \bibinfo{journal}{Automatica}\/},  {\it \bibinfo{volume}{45}\/},
  \bibinfo{pages}{1771--1777}.
\bibitem[{Calafiore \& Campi(2006)}]{GCC-MCC:06}
\bibinfo{author}{Calafiore, G.~C.}, \& \bibinfo{author}{Campi, M.~C.}
  (\bibinfo{year}{2006}).
\newblock \bibinfo{title}{The scenario approach to robust control design}.
\newblock {\it \bibinfo{journal}{IEEE Transactions on Automatic Control}\/},
  {\it \bibinfo{volume}{51}\/}, \bibinfo{pages}{742--753}.
\bibitem[{Cardaliaguet(1996)}]{PC:96}
\bibinfo{author}{Cardaliaguet, P.} (\bibinfo{year}{1996}).
\newblock \bibinfo{title}{A differential game with two players and one target}.
\newblock {\it \bibinfo{journal}{SIAM Journal on Control and Optimization}\/},
  {\it \bibinfo{volume}{34}\/}, \bibinfo{pages}{1441--1460}.
\bibitem[{{Chen} et~al.(2019){Chen}, {Bansal}, {Fisac} \&
  {Tomlin}}]{MC-SB-JFF-CJT:19}
\bibinfo{author}{{Chen}, M.}, \bibinfo{author}{{Bansal}, S.},
  \bibinfo{author}{{Fisac}, J.~F.}, \& \bibinfo{author}{{Tomlin}, C.~J.}
  (\bibinfo{year}{2019}).
\newblock \bibinfo{title}{Robust sequential trajectory planning under
  disturbances and adversarial intruder}.
\newblock {\it \bibinfo{journal}{IEEE Transactions on Control Systems
  Technology}\/},  {\it \bibinfo{volume}{27}\/}, \bibinfo{pages}{1566--1582}.
\bibitem[{{Chen} et~al.(2018){Chen}, {Herbert}, {Vashishtha}, {Bansal} \&
  {Tomlin}}]{MC-SLH-MSV-SB-CJT:18}
\bibinfo{author}{{Chen}, M.}, \bibinfo{author}{{Herbert}, S.~L.},
  \bibinfo{author}{{Vashishtha}, M.~S.}, \bibinfo{author}{{Bansal}, S.}, \&
  \bibinfo{author}{{Tomlin}, C.~J.} (\bibinfo{year}{2018}).
\newblock \bibinfo{title}{Decomposition of reachable sets and tubes for a class
  of nonlinear systems}.
\newblock {\it \bibinfo{journal}{IEEE Transactions on Automatic Control}\/},
  {\it \bibinfo{volume}{63}\/}, \bibinfo{pages}{3675--3688}.
\bibitem[{{Chen} et~al.(2017){Chen}, {Zhou} \& {Tomlin}}]{MC-ZZ-CJT:17}
\bibinfo{author}{{Chen}, M.}, \bibinfo{author}{{Zhou}, Z.}, \&
  \bibinfo{author}{{Tomlin}, C.~J.} (\bibinfo{year}{2017}).
\newblock \bibinfo{title}{Multiplayer reach-avoid games via pairwise outcomes}.
\newblock {\it \bibinfo{journal}{IEEE Transactions on Automatic Control}\/},
  {\it \bibinfo{volume}{62}\/}, \bibinfo{pages}{1451--1457}.
\bibitem[{Elliott \& Kalton(1972)}]{RJE-NJK:72}
\bibinfo{author}{Elliott, R.~J.}, \& \bibinfo{author}{Kalton, N.~J.}
  (\bibinfo{year}{1972}).
\newblock {\it \bibinfo{title}{The existence of value in differential
  games}\/}.
\newblock \bibinfo{publisher}{American Mathematical Soc.}
\bibitem[{Exarchos et~al.(2015)Exarchos, Tsiotras \& Pachter}]{IE-PT-MP:15}
\bibinfo{author}{Exarchos, I.}, \bibinfo{author}{Tsiotras, P.}, \&
  \bibinfo{author}{Pachter, M.} (\bibinfo{year}{2015}).
\newblock \bibinfo{title}{On the suicidal pedestrian differential game}.
\newblock {\it \bibinfo{journal}{Dynamic Games and Applications}\/},  {\it
  \bibinfo{volume}{5}\/}, \bibinfo{pages}{297--317}.
\bibitem[{Falcone(2006)}]{MF:06}
\bibinfo{author}{Falcone, M.} (\bibinfo{year}{2006}).
\newblock \bibinfo{title}{Numerical methods for differential games based on
  partial differential equations}.
\newblock {\it \bibinfo{journal}{International Game Theory Review}\/},  {\it
  \bibinfo{volume}{8}\/}, \bibinfo{pages}{231--272}.
\bibitem[{Fonod \& Shima(2018)}]{RF-TS:18}
\bibinfo{author}{Fonod, R.}, \& \bibinfo{author}{Shima, T.}
  (\bibinfo{year}{2018}).
\newblock \bibinfo{title}{Blinding guidance against missiles sharing
  bearings-only measurements}.
\newblock {\it \bibinfo{journal}{IEEE Transactions on Aerospace and Electronic
  Systems}\/},  {\it \bibinfo{volume}{54}\/}, \bibinfo{pages}{205--216}.
\bibitem[{Fu \& Liu(2023)}]{HF-HHL:23}
\bibinfo{author}{Fu, H.}, \& \bibinfo{author}{Liu, H. H.-T.}
  (\bibinfo{year}{2023}).
\newblock \bibinfo{title}{Justification of the geometric solution of a target
  defense game with faster defenders and a convex target area using the {HJI}
  equation}.
\newblock {\it \bibinfo{journal}{Automatica}\/},  {\it
  \bibinfo{volume}{149}\/}, \bibinfo{pages}{110811}.
\bibitem[{Garcia et~al.(2020{\natexlab{a}})Garcia, Casbeer \&
  Pachter}]{EG-DWC-MP:20}
\bibinfo{author}{Garcia, E.}, \bibinfo{author}{Casbeer, D.~W.}, \&
  \bibinfo{author}{Pachter, M.} (\bibinfo{year}{2020}{\natexlab{a}}).
\newblock \bibinfo{title}{Optimal strategies for a class of multi-player
  reach-avoid differential games in {3D} space}.
\newblock {\it \bibinfo{journal}{IEEE Robotics and Automation Letters}\/},
  {\it \bibinfo{volume}{5}\/}, \bibinfo{pages}{4257--4264}.
\bibitem[{Garcia et~al.(2020{\natexlab{b}})Garcia, Casbeer, Von~Moll \&
  Pachter}]{EG-DWS-AVM-MP:20}
\bibinfo{author}{Garcia, E.}, \bibinfo{author}{Casbeer, D.~W.},
  \bibinfo{author}{Von~Moll, A.}, \& \bibinfo{author}{Pachter, M.}
  (\bibinfo{year}{2020}{\natexlab{b}}).
\newblock \bibinfo{title}{Multiple pursuer multiple evader differential games}.
\newblock {\it \bibinfo{journal}{IEEE Transactions on Automatic Control}\/},
  {\it \bibinfo{volume}{66}\/}, \bibinfo{pages}{2345--2350}.
\bibitem[{Getz \& Pachter(1981)}]{WMG-MP:81}
\bibinfo{author}{Getz, W.}, \& \bibinfo{author}{Pachter, M.}
  (\bibinfo{year}{1981}).
\newblock \bibinfo{title}{Two-target pursuit-evasion differential games in the
  plane}.
\newblock {\it \bibinfo{journal}{Journal of Optimization Theory and
  Applications}\/},  {\it \bibinfo{volume}{34}\/}, \bibinfo{pages}{383--403}.
\bibitem[{Isaacs(1965)}]{RI:65}
\bibinfo{author}{Isaacs, R.} (\bibinfo{year}{1965}).
\newblock {\it \bibinfo{title}{Differential Games}\/}.
\newblock \bibinfo{publisher}{New York: Wiley}.
\bibitem[{Karp(1972)}]{RMK:72}
\bibinfo{author}{Karp, R.~M.} (\bibinfo{year}{1972}).
\newblock \bibinfo{title}{Reducibility among combinatorial problems}.
\newblock In {\it \bibinfo{booktitle}{Complexity of computer computations}\/}
  (pp. \bibinfo{pages}{85--103}).
\bibitem[{Lee \& Bakolas(2022{\natexlab{a}})}]{YL-EB:22-2}
\bibinfo{author}{Lee, Y.}, \& \bibinfo{author}{Bakolas, E.}
  (\bibinfo{year}{2022}{\natexlab{a}}).
\newblock \bibinfo{title}{Guarding a convex target set from an attacker in
  {E}uclidean spaces}.
\newblock {\it \bibinfo{journal}{IEEE Control Systems Letters}\/},  {\it
  \bibinfo{volume}{6}\/}, \bibinfo{pages}{1706--1711}.
\bibitem[{Lee \& Bakolas(2022{\natexlab{b}})}]{YL-EB:22}
\bibinfo{author}{Lee, Y.}, \& \bibinfo{author}{Bakolas, E.}
  (\bibinfo{year}{2022}{\natexlab{b}}).
\newblock \bibinfo{title}{Solutions for games of guarding an arbitrary convex
  target set with multiple defenders in ${R}^n$}.
\newblock {\it \bibinfo{journal}{TechRxiv: dx.doi.org/10.36227
  /techrxiv.20468970.v1}\/}, .
\bibitem[{Lewin \& Breakwell(1975)}]{JL-JVB:75}
\bibinfo{author}{Lewin, J.}, \& \bibinfo{author}{Breakwell, J.}
  (\bibinfo{year}{1975}).
\newblock \bibinfo{title}{The surveillance-evasion game of degree}.
\newblock {\it \bibinfo{journal}{Journal of Optimization Theory and
  Applications}\/},  {\it \bibinfo{volume}{16}\/}, \bibinfo{pages}{339--353}.
\bibitem[{Lewin \& Olsder(1979)}]{JL-GJO-79}
\bibinfo{author}{Lewin, J.}, \& \bibinfo{author}{Olsder, G.}
  (\bibinfo{year}{1979}).
\newblock \bibinfo{title}{Conic surveillance evasion}.
\newblock {\it \bibinfo{journal}{Journal of Optimization Theory and
  Applications}\/},  {\it \bibinfo{volume}{27}\/}, \bibinfo{pages}{107--125}.
\bibitem[{Li et~al.(2005)Li, Cruz, Chen, Kwan \& Chang}]{DL-JBC-GC-CK-MHC:05}
\bibinfo{author}{Li, D.}, \bibinfo{author}{Cruz, J.~B.}, \bibinfo{author}{Chen,
  G.}, \bibinfo{author}{Kwan, C.}, \& \bibinfo{author}{Chang, M.-H.}
  (\bibinfo{year}{2005}).
\newblock \bibinfo{title}{A hierarchical approach to multi-player
  pursuit-evasion differential games}.
\newblock In {\it \bibinfo{booktitle}{Proceedings of the 44th IEEE Conference
  on Decision and Control}\/} (pp. \bibinfo{pages}{5674--5679}).
\bibitem[{Makkapati \& Tsiotras(2019)}]{VRM-PT:19}
\bibinfo{author}{Makkapati, V.~R.}, \& \bibinfo{author}{Tsiotras, P.}
  (\bibinfo{year}{2019}).
\newblock \bibinfo{title}{Optimal evading strategies and task allocation in
  multi-player pursuit--evasion problems}.
\newblock {\it \bibinfo{journal}{Dynamic Games and Applications}\/},  {\it
  \bibinfo{volume}{9}\/}, \bibinfo{pages}{1168--1187}.
\bibitem[{{Margellos} \& {Lygeros}(2011)}]{KM-JL:11}
\bibinfo{author}{{Margellos}, K.}, \& \bibinfo{author}{{Lygeros}, J.}
  (\bibinfo{year}{2011}).
\newblock \bibinfo{title}{Hamilton–{J}acobi formulation for reach–avoid
  differential games}.
\newblock {\it \bibinfo{journal}{IEEE Transactions on Automatic Control}\/},
  {\it \bibinfo{volume}{56}\/}, \bibinfo{pages}{1849--1861}.
\bibitem[{Merz(1971)}]{AWM:71}
\bibinfo{author}{Merz, A.~W.} (\bibinfo{year}{1971}).
\newblock {\it \bibinfo{title}{The homicidal chauffeur--a differential
  game}\/}.
\newblock Ph.D. thesis Stanford University.
\bibitem[{Merz(1974)}]{AWM:74}
\bibinfo{author}{Merz, A.~W.} (\bibinfo{year}{1974}).
\newblock \bibinfo{title}{The homicidal chauffeur}.
\newblock {\it \bibinfo{journal}{AIAA Journal}\/},  {\it
  \bibinfo{volume}{12}\/}, \bibinfo{pages}{259--260}.
\bibitem[{Mitchell(2002)}]{IMM:02}
\bibinfo{author}{Mitchell, I.~M.} (\bibinfo{year}{2002}).
\newblock {\it \bibinfo{title}{Application of level set methods to control and
  reachability problems in continuous and hybrid systems}\/}.
\newblock Ph.D. thesis Stanford University.
\bibitem[{{Mitchell} et~al.(2005){Mitchell}, {Bayen} \&
  {Tomlin}}]{IMM-AMB-CJT:05}
\bibinfo{author}{{Mitchell}, I.~M.}, \bibinfo{author}{{Bayen}, A.~M.}, \&
  \bibinfo{author}{{Tomlin}, C.~J.} (\bibinfo{year}{2005}).
\newblock \bibinfo{title}{A time-dependent {H}amilton-{J}acobi formulation of
  reachable sets for continuous dynamic games}.
\newblock {\it \bibinfo{journal}{IEEE Transactions on Automatic Control}\/},
  {\it \bibinfo{volume}{50}\/}, \bibinfo{pages}{947--957}.
\bibitem[{Mohanan et~al.(2020)Mohanan, Kothuri \& Bhikkaji}]{JM-NK-BB:20}
\bibinfo{author}{Mohanan, J.}, \bibinfo{author}{Kothuri, N.}, \&
  \bibinfo{author}{Bhikkaji, B.} (\bibinfo{year}{2020}).
\newblock \bibinfo{title}{The target guarding problem: A real time solution for
  noise corrupted measurements}.
\newblock {\it \bibinfo{journal}{European Journal of Control}\/},  {\it
  \bibinfo{volume}{54}\/}, \bibinfo{pages}{111--118}.
\bibitem[{Oyler et~al.(2016)Oyler, Kabamba \& Girard}]{DWO-PTK-ARG:16}
\bibinfo{author}{Oyler, D.~W.}, \bibinfo{author}{Kabamba, P.~T.}, \&
  \bibinfo{author}{Girard, A.~R.} (\bibinfo{year}{2016}).
\newblock \bibinfo{title}{Pursuit--evasion games in the presence of obstacles}.
\newblock {\it \bibinfo{journal}{Automatica}\/},  {\it \bibinfo{volume}{65}\/},
  \bibinfo{pages}{1--11}.
\bibitem[{Pachter \& Yavin(1981)}]{MP-YY:81}
\bibinfo{author}{Pachter, M.}, \& \bibinfo{author}{Yavin, Y.}
  (\bibinfo{year}{1981}).
\newblock \bibinfo{title}{A stochastic homicidal chauffeur pursuit-evasion
  differential game}.
\newblock {\it \bibinfo{journal}{Journal of Optimization Theory and
  Applications}\/},  {\it \bibinfo{volume}{34}\/}, \bibinfo{pages}{405--424}.
\bibitem[{Patsko \& Turova(2011)}]{VSP-VLT:11}
\bibinfo{author}{Patsko, V.~S.}, \& \bibinfo{author}{Turova, V.~L.}
  (\bibinfo{year}{2011}).
\newblock \bibinfo{title}{Homicidal {C}hauffeur game: History and modern
  studies}.
\newblock In {\it \bibinfo{booktitle}{Advances in Dynamic Games: Theory,
  Applications, and Numerical Methods for Differential and Stochastic Games}\/}
  (pp. \bibinfo{pages}{227--251}).
\bibitem[{Pihl et~al.(2012)Pihl, Marcher \& Jensen}]{pihl2012phased}
\bibinfo{author}{Pihl, M.~J.}, \bibinfo{author}{Marcher, J.}, \&
  \bibinfo{author}{Jensen, J.~A.} (\bibinfo{year}{2012}).
\newblock \bibinfo{title}{Phased-array vector velocity estimation using
  transverse oscillations}.
\newblock {\it \bibinfo{journal}{IEEE Transactions on Ultrasonics,
  Ferroelectrics, and Frequency Control}\/},  {\it \bibinfo{volume}{59}\/},
  \bibinfo{pages}{2662--2675}.
\bibitem[{Shishika \& Kumar(2018)}]{DS-VK:18}
\bibinfo{author}{Shishika, D.}, \& \bibinfo{author}{Kumar, V.}
  (\bibinfo{year}{2018}).
\newblock \bibinfo{title}{Local-game decomposition for multiplayer
  perimeter-defense problem}.
\newblock In {\it \bibinfo{booktitle}{IEEE Conference on Decision and
  Control}\/} (pp. \bibinfo{pages}{2093--2100}).
\bibitem[{Shishika \& Kumar(2020)}]{DS-VK:20}
\bibinfo{author}{Shishika, D.}, \& \bibinfo{author}{Kumar, V.}
  (\bibinfo{year}{2020}).
\newblock \bibinfo{title}{A review of multi agent perimeter defense games}.
\newblock In {\it \bibinfo{booktitle}{International Conference on Decision and
  Game Theory for Security}\/} (pp. \bibinfo{pages}{472--485}).
\bibitem[{Soumekh(1997)}]{soumekh1997phased}
\bibinfo{author}{Soumekh, M.} (\bibinfo{year}{1997}).
\newblock \bibinfo{title}{Phased array imaging of moving targets with
  randomized beam steering and area spotlighting}.
\newblock {\it \bibinfo{journal}{IEEE Transactions on Image Processing}\/},
  {\it \bibinfo{volume}{6}\/}, \bibinfo{pages}{736--749}.
\bibitem[{Von~Moll et~al.(2020)Von~Moll, Garcia, Casbeer, Suresh \&
  Swar}]{AVM-EG-DC-MS-SCS:20}
\bibinfo{author}{Von~Moll, A.}, \bibinfo{author}{Garcia, E.},
  \bibinfo{author}{Casbeer, D.}, \bibinfo{author}{Suresh, M.}, \&
  \bibinfo{author}{Swar, S.~C.} (\bibinfo{year}{2020}).
\newblock \bibinfo{title}{Multiple-pursuer, single-evader border defense
  differential game}.
\newblock {\it \bibinfo{journal}{Journal of Aerospace Information Systems}\/},
  {\it \bibinfo{volume}{17}\/}, \bibinfo{pages}{407--416}.
\bibitem[{Von~Moll et~al.(2022)Von~Moll, Pachter, Shishika \&
  Fuchs}]{AVM-MP-DS-ZF:22}
\bibinfo{author}{Von~Moll, A.}, \bibinfo{author}{Pachter, M.},
  \bibinfo{author}{Shishika, D.}, \& \bibinfo{author}{Fuchs, Z.}
  (\bibinfo{year}{2022}).
\newblock \bibinfo{title}{Circular target defense differential games$^*$}.
\newblock {\it \bibinfo{journal}{IEEE Transactions on Automatic Control}\/},
  {\it \bibinfo{volume}{1}\/}, \bibinfo{pages}{1--14}.
\bibitem[{Wasz et~al.(2019)Wasz, Pachter \& Pham}]{PW-MP-KP:19}
\bibinfo{author}{Wasz, P.}, \bibinfo{author}{Pachter, M.}, \&
  \bibinfo{author}{Pham, K.} (\bibinfo{year}{2019}).
\newblock \bibinfo{title}{Two-an-one pursuit with a non-zero capture radius}.
\newblock In {\it \bibinfo{booktitle}{2019 27th Mediterranean Conference on
  Control and Automation (MED)}\/} (pp. \bibinfo{pages}{577--582}).
\bibitem[{Yan et~al.(2023{\natexlab{a}})Yan, Deng, Duan, Shi \&
  Zhong}]{RY-RD-XD-ZS-YZ:23}
\bibinfo{author}{Yan, R.}, \bibinfo{author}{Deng, R.}, \bibinfo{author}{Duan,
  X.}, \bibinfo{author}{Shi, Z.}, \& \bibinfo{author}{Zhong, Y.}
  (\bibinfo{year}{2023}{\natexlab{a}}).
\newblock \bibinfo{title}{Multiplayer reach-avoid differential games with
  simple motions: A review}.
\newblock {\it \bibinfo{journal}{Frontiers in Control Engineering}\/},  {\it
  \bibinfo{volume}{3}\/}.
\bibitem[{Yan et~al.(2023{\natexlab{b}})Yan, Deng, Lai, Zhang, Shi \&
  Zhong}]{RY-RD-HL-WZ-ZS-YZ:23}
\bibinfo{author}{Yan, R.}, \bibinfo{author}{Deng, R.}, \bibinfo{author}{Lai,
  H.}, \bibinfo{author}{Zhang, W.}, \bibinfo{author}{Shi, Z.}, \&
  \bibinfo{author}{Zhong, Y.} (\bibinfo{year}{2023}{\natexlab{b}}).
\newblock \bibinfo{title}{Homicidal {C}hauffeur reach-avoid games via
  guaranteed winning strategies}.
\newblock {\it \bibinfo{journal}{IEEE Transactions on Automatic Control}\/},
  (pp. \bibinfo{pages}{1--16}).
\bibitem[{Yan et~al.(2022)Yan, Duan, Shi, Zhong \& Bullo}]{RY-XD-ZS-YZ-FB:19}
\bibinfo{author}{Yan, R.}, \bibinfo{author}{Duan, X.}, \bibinfo{author}{Shi,
  Z.}, \bibinfo{author}{Zhong, Y.}, \& \bibinfo{author}{Bullo, F.}
  (\bibinfo{year}{2022}).
\newblock \bibinfo{title}{Matching-based capture strategies for {3D}
  heterogeneous multiplayer reach-avoid differential games}.
\newblock {\it \bibinfo{journal}{Automatica}\/},  {\it
  \bibinfo{volume}{140}\/}, \bibinfo{pages}{110207}.
\bibitem[{{Yan} et~al.(2019){Yan}, {Shi} \& {Zhong}}]{RY-ZS-YZ:19}
\bibinfo{author}{{Yan}, R.}, \bibinfo{author}{{Shi}, Z.}, \&
  \bibinfo{author}{{Zhong}, Y.} (\bibinfo{year}{2019}).
\newblock \bibinfo{title}{Reach-avoid games with two defenders and one
  attacker: An analytical approach}.
\newblock {\it \bibinfo{journal}{IEEE Transactions on Cybernetics}\/},  {\it
  \bibinfo{volume}{49}\/}, \bibinfo{pages}{1035--1046}.
\bibitem[{Yan et~al.(2020)Yan, Shi \& Zhong}]{RY-ZS-YZ:20-2}
\bibinfo{author}{Yan, R.}, \bibinfo{author}{Shi, Z.}, \&
  \bibinfo{author}{Zhong, Y.} (\bibinfo{year}{2020}).
\newblock \bibinfo{title}{Guarding a subspace in high-dimensional space with
  two defenders and one attacker}.
\newblock {\it \bibinfo{journal}{IEEE Transactions on Cybernetics}\/},  {\it
  \bibinfo{volume}{52}\/}, \bibinfo{pages}{3998--4011}.
\bibitem[{{Yan} et~al.(2020){Yan}, {Shi} \& {Zhong}}]{RY-ZS-YZ:20-1}
\bibinfo{author}{{Yan}, R.}, \bibinfo{author}{{Shi}, Z.}, \&
  \bibinfo{author}{{Zhong}, Y.} (\bibinfo{year}{2020}).
\newblock \bibinfo{title}{Task assignment for multiplayer reach-avoid games in
  convex domains via analytical barriers}.
\newblock {\it \bibinfo{journal}{IEEE Transactions on Robotics}\/},  {\it
  \bibinfo{volume}{36}\/}, \bibinfo{pages}{107--124}.
\bibitem[{Yan et~al.(2021)Yan, Shi \& Zhong}]{RY-ZS-YZ:19-2}
\bibinfo{author}{Yan, R.}, \bibinfo{author}{Shi, Z.}, \&
  \bibinfo{author}{Zhong, Y.} (\bibinfo{year}{2021}).
\newblock \bibinfo{title}{Optimal strategies for the lifeline differential game
  with limited lifetime}.
\newblock {\it \bibinfo{journal}{International Journal of Control}\/},  {\it
  \bibinfo{volume}{94}\/}, \bibinfo{pages}{2238--2251}.
\bibitem[{{Zhou} et~al.(2012){Zhou}, {Takei}, {Huang} \&
  {Tomlin}}]{ZZ-RT-HH-CJT:12}
\bibinfo{author}{{Zhou}, Z.}, \bibinfo{author}{{Takei}, R.},
  \bibinfo{author}{{Huang}, H.}, \& \bibinfo{author}{{Tomlin}, C.~J.}
  (\bibinfo{year}{2012}).
\newblock \bibinfo{title}{A general, open-loop formulation for reach-avoid
  games}.
\newblock In {\it \bibinfo{booktitle}{IEEE Conference on Decision and
  Control}\/} (pp. \bibinfo{pages}{6501--6506}).
\bibitem[{Zhou et~al.(2016)Zhou, Zhang, Ding, Huang, Stipanovi{\'c} \&
  Tomlin}]{ZZ-WZ-JD-HH-DMS-CJT:16}
\bibinfo{author}{Zhou, Z.}, \bibinfo{author}{Zhang, W.}, \bibinfo{author}{Ding,
  J.}, \bibinfo{author}{Huang, H.}, \bibinfo{author}{Stipanovi{\'c}, D.~M.}, \&
  \bibinfo{author}{Tomlin, C.~J.} (\bibinfo{year}{2016}).
\newblock \bibinfo{title}{Cooperative pursuit with voronoi partitions}.
\newblock {\it \bibinfo{journal}{Automatica}\/},  {\it \bibinfo{volume}{72}\/},
  \bibinfo{pages}{64--72}.

\end{thebibliography}



\end{document}